\documentclass[12pt]{iopart}
\usepackage{epsfig}
\begin{document}

\title{Magnetospectroscopy of symmetric and anti-symmetric states in double quantum wells}

\author{M. Marchewka, E. M. Sheregii, I. Tralle, D. Ploch, G. Tomaka, M. Furdak}
\address{Institute of Physics, University of Rzesz\'ow,\\ Rejtana 16a,
35-310 Rzesz\'ow, Poland.}
\author{A. Kolek, A. Stadler, K. Mleczko, D. Zak}
\address{Department of Fundamental Electronics Rzesz\'ow University of Technology,
 35-959 Rzeszów, W. Pola 2, Poland.}
\author{W. Strupinski, A. Jasik, R. Jakiela}
\address{Institute of Electronic Materials Technology, 
W\'olczyñska 133, 01-919 Warsaw, Poland.}

\ead{marmi@univ.rzeszow.pl}
\begin{abstract}
The experimental results obtained for the magneto-transport in the InGaAs/InAlAs double quantum wells (DQW) structures of two different shapes of wells are reported. The beating-effect occurred in the Shubnikov-de Haas (SdH) oscillations was observed for both types of the structures at low temperatures in the parallel transport when magnetic field was perpendicular to the layers. An approach to the calculation of the Landau levels energies for DQW structures was developed and then applied to the analysis and interpretation of the experimental data related to the beating-effect. We also argue that in order to account for the observed magneto-transport phenomena (SdH and Integer Quantum Hall effect), one should introduce two different quasi-Fermi levels characterizing two electron sub-systems regarding symmetry properties of their states, symmetric and anti-symmetric ones which are not mixed by electron-electron interaction. 
\end{abstract}

\maketitle

\section{Introduction}

In recent years a great number of publications appeared, where the experimental studies of electron transport in Double Quantum Well  (DQW) structures were reported. The reason for this is the enduring progress in the structure fabrication technology which allows to modify the shape and width of QW or barriers, as well as the progress in measurement techniques enables to measure the separate conductivity and Hall response of each layer. These experiments have demonstrated a number of new and interesting phenomena occurring in such structures which however, cannot be observed in a structure with a single two-dimensional-gas (2DEG) layer. Generally, all these phenomena can be divided into two groups depending on the intensity of interlayer interaction. To the first group the phenomena related to the Coulomb drag effect belong, while to the second one, the phenomena related to the tunneling effect. Correspondently,  the phenomena of the first group are accounted for the weak interaction between the layers, while the phenomena of the second group are attributed to the strong interaction.

In order to reveal more clearly drag effect caused by Coulomb interaction, the tunneling between two QWs has to be negligible and in order to fulfill this requirement, the barrier between two QWs should be adequately wide. M.C. Bronsager \textit{et. }al\cite{1} reported theoretical results concerning the magnetic-field and temperature dependences of the trans-resistivity in DQW-layers. Experiments of Ref. 2 confirmed the results obtained in Ref. 1 and have shown also the existence of double-peak structure for each magneto-resistance peak in the samples characterized by the matched density of states in the layers. Another type of experiments were carried out on the samples with unmatched density of states\cite{3}. In this work authors demonstrated the existence of negative drag in the IQHE-regime; its clear explanation was proposed by J. G. S. Lok \textit{et. }al\cite{4}. The authors of Ref. 4 argued that the electron spin plays decisive role in the drag effect. At last, the fractional drag between parallel two-dimensional systems has been observed and measured in a regime of strong interlayer correlations\cite{5}, where the Hall drag resistance was observed to be quantized at $\hbar /e^{2} $. 

Another situation takes place if one has two non-identical QWs. In this case some of the integer Hall states appear and some others disappear, which depends on alignment of electron states in two QWs\cite{6}.

 A lot of new phenomena caused by the tunneling between two 2DEG-layers separated by the barrier were also observed in recent years. One of them is the absence of some integer quantum Hall states caused by the splitting of electron states into symmetric and anti-symmetric ones (so colled SAS-gap, or $\Delta _{SAS} $) in DQW systems\cite{7,8}.

 In a structure where the barrier is thin enough, the SAS-gap plays decisive role. The magnitude of this splitting can be determined directly by means of current-voltage characteristics of such structure; it was proven also that the SAS-gap is proportional to the magnetic field\cite{9}. This effect was attributed by D. Huang and M.O. Manasreh\cite{10} to the screening of electron-electron interaction in the DQW-structure.

 In a last few years many new phenomena were observed in the structures with strongly correlated 2DEG at very low temperatures, among them the Bose condensation\cite{11}, Wigner crystallization\cite{12,13}, and the long-time nuclear-lattice relaxation in DQW\cite{14,15}. Some authors\cite{16,17} have considered the Shubnikov-de Haas (SdH) oscillations in a tilted magnetic field with the strong perpendicular $B_{\bot } $ and small parallel $B_{||}$ components of the field. The authors of Ref. 17 stated that the small $B_{||} $- component is responsible for the beating effect in SdH oscillations occurred in DQW-structure. If there is no $B_{\bot } $-component and magnetic field is parallel to the layers composing DQW, the most important peculiarities of magneto-transport are accounted for the crossing of Fermi-surfaces belonging to two QWs\cite{18,19,20} and this case is completely different from two previous ones. 

Therefore, as for the electrons in strongly correlated DQW-structures is concerned, it is possible to treat them as two separate sub-systems, one of them as belonging to symmetric states, while another one as to belonging to the anti-symmetric ones. The symmetric and anti-symmetric states differ from one another because the symmetric properties of their wave functions are different. To the best of our knowledge, in the papers published by far, except of our short communications\cite{21,22}, the consequences of this essential difference between the symmetric properties of the electron's wave functions were not treated properly. It is worth to be mentioned also, that the magnetotransport phenomena in such structures were not analyzed previously by taking into account the accurately calculated energies of Landau levels (LL).

 In this paper the parallel magneto-transport in DQW-structures composed of two identical QWs separated by thin barrier is studied. The experimental results are obtained for two different structures, the first one included QWs of quasi-rectangular shape, while the second one included the triangular QWs. Both of the structures are based on the InGaAs/InAlAs-system and for both of them we observed the beating effect in SdH-oscillations in perpendicular magnetic field arrangement (that is, without $B_{||} $ -component of the field). As it will be seen below, in order to interpret the beating effect which occurs in SdH-oscillations in the DQW with the matched density of states and in the absence of $B_{||} $-component of magnetic field, it is necessary to introduce two different quasi-Fermi levels for the symmetric and anti-symmetric electron sub-systems.

\section{Description of the structures}

 The DQW based on InGaAs/InAlAs/InP structures were produced by means of the low pressure metal organic vapour phase epitaxy (LP-MOVPE) on semi-insulating (100) InP: Fe substrates at the Institute of Electronic Materials Technology, Warsaw. The method of producing the structures with a single QW, which was reported earlier\cite{23,24}, was used also for producing DQW structure with rectangular, as well as triangular QWs. The structure of the first type (\#2506) consists of two  In$_{0.65}$Ga$_{0.35}$As QW of about 20 nm thickness each, and three In$_{0.52}$Al$_{0.48}$As barriers. In each barrier there was the donor $\delta $-doping layer (see Fig. 1).

\begin{table}[h]
\caption{\label{label}Parameters of the InGaAs/InAlAs DQW structure.}
\begin{tabular}{@{}llllll}
\br
           &Concentration       &Thickness         &               &Number of           & $\delta$-doping layers\\
Sample      &of (\%) InAs in    &   of the         &QW-profile     &  $\delta$-doping   & with donor 																																																				concentration\\
            &the channels       &channels          &               &   layers           & (according to 																																																		technologists) \\
\mr
\#2506      & 65                & 20 nm          & smooth interface & 3               & 3.5 $\times 10^{12} cm^{-2}$\\
\#3181      & 65                & 5 nm          & sharp interface  & 2                & 3.5 $\times 10^{12} cm^{-2}$\\
\br
\end{tabular}
\end{table}

\begin{figure}[ht]
\begin{indented}
\item[]
\includegraphics[scale=0.65]{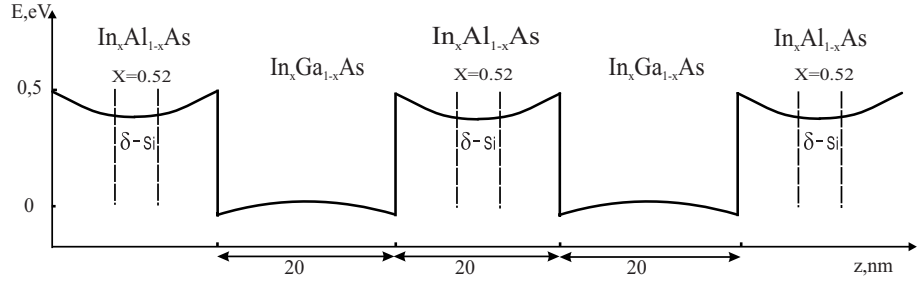}
\caption{\label{fig:epsart} The conduction band edge profile of the
\#2506 structure.}
\end{indented}
\end{figure}
\begin{figure}[ht]
\begin{indented}
\item[]
\includegraphics[scale=0.30] {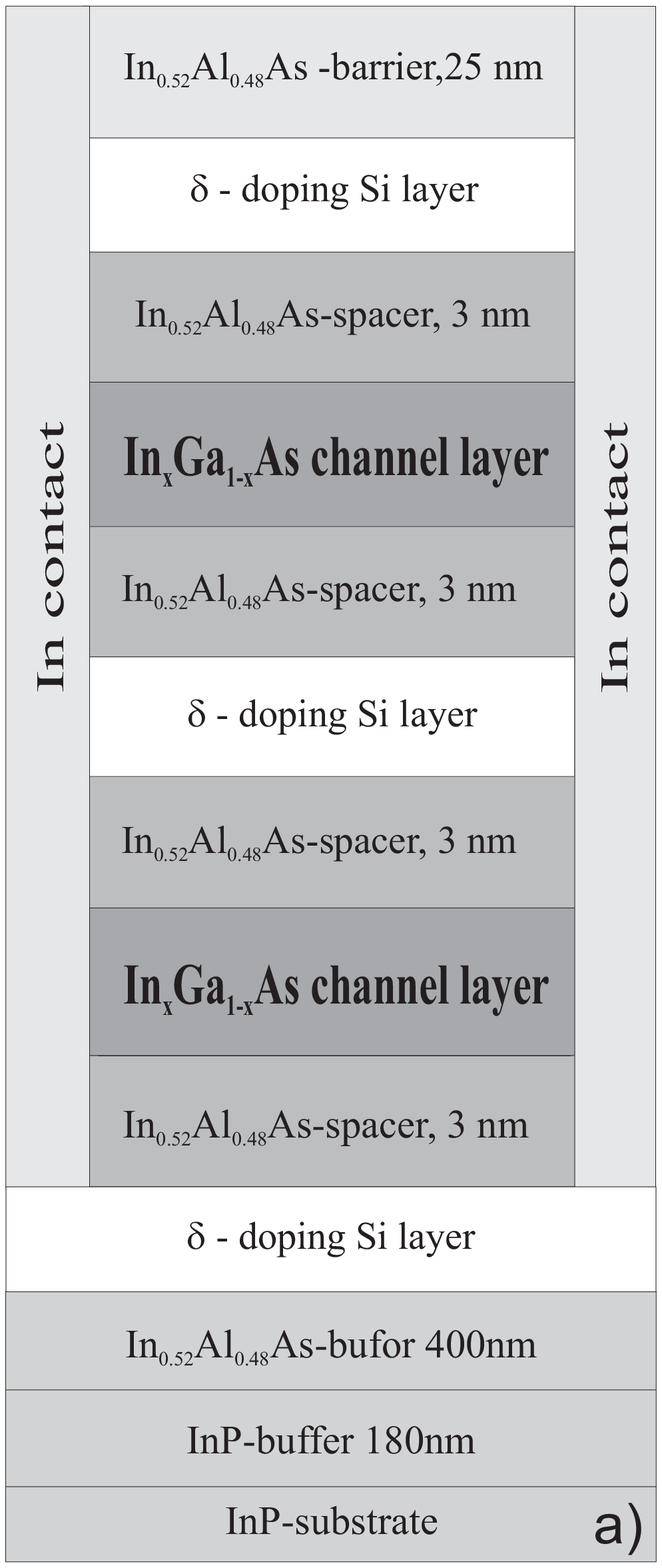}
\includegraphics[scale=0.38] {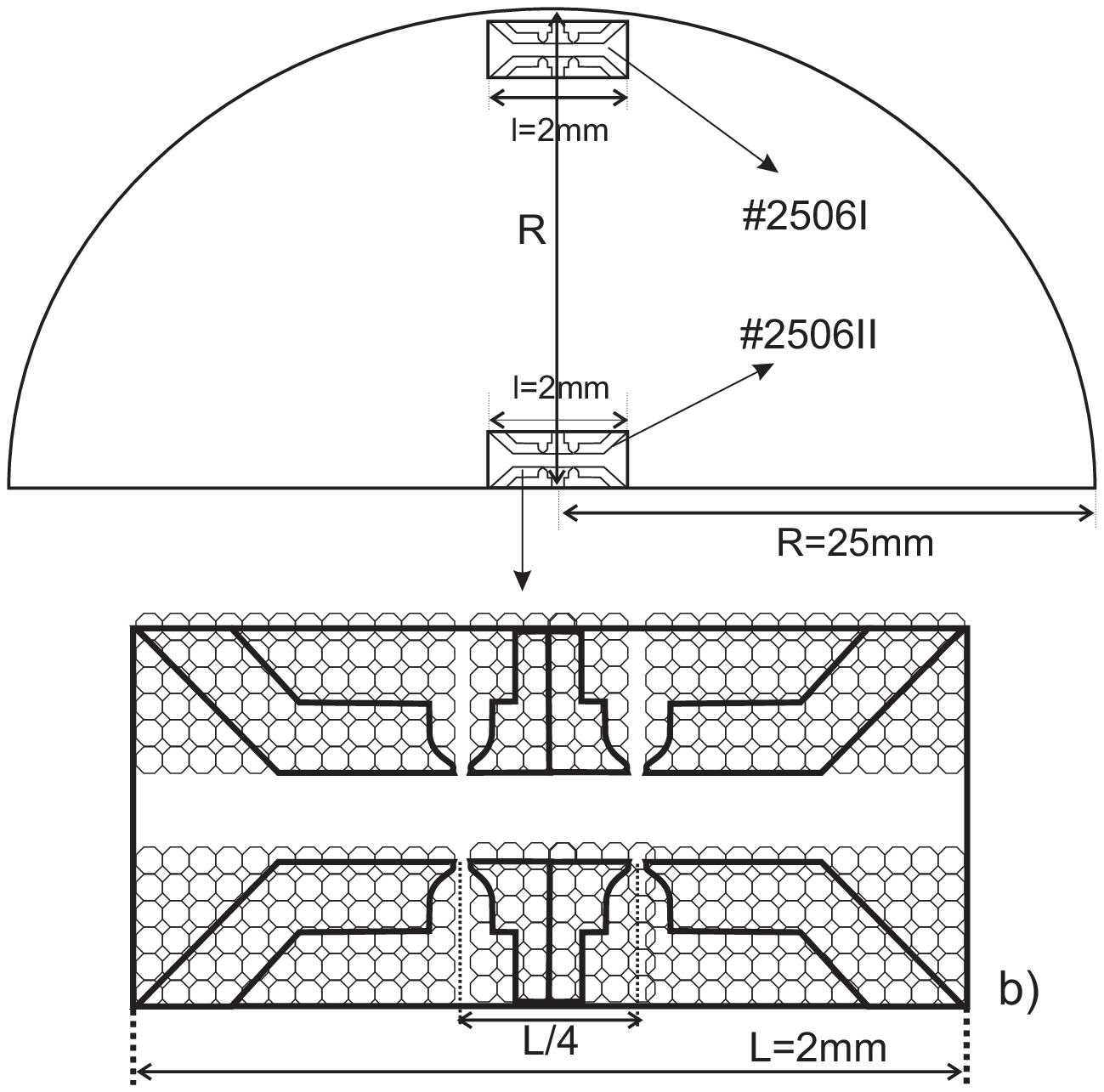}
\includegraphics[scale=0.30] {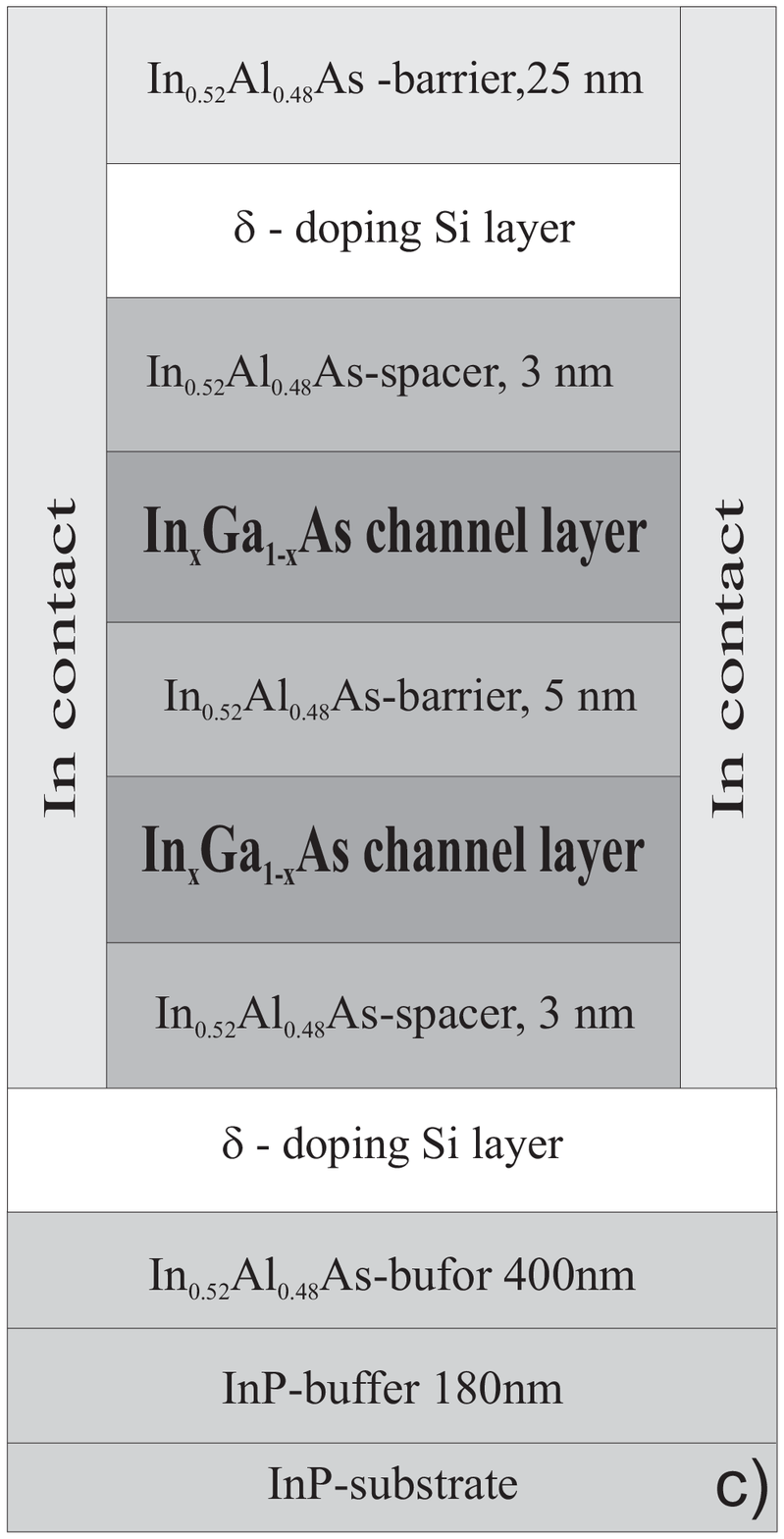}
\caption{\label{fig:epsart}a) The cross-section of the \#2506 structure used in the experiments; b) upper one: the piece of wafer with the indicated parts of it from where the \#2506I and \#2506II was cut out; lower one: view of the sample; c) The cross-section of the \#3181 structure.}
\end{indented}
\end{figure}

The cross section of \#2506-structure is depicted in Fig. 2a. The second DQW structure (\#3181) whose cross-section is shown in Fig. 2c, also consists of two In$_{0.65}$Ga$_{0.35}$As QW of the thickness of about 5 nm and three In$_{0.52}$Al$_{0.48}$As barriers. In this case however, the barrier between QWs was not doped. The corresponding conduction band shape is shown in Fig. 3. For this particular case the 2DEG in each QW could be considered as to be putted into a triangular potential well, since as it is shown in Fig. 3, the bottom of the conduction band in each QW makes a triangle with the horizontal line.

\begin{figure}[ht]
\begin{indented}
\item[]
\includegraphics[scale=0.7]{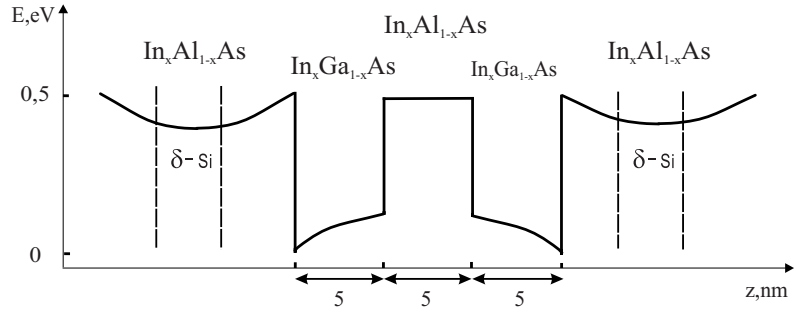}
\caption{\label{fig:epsart}The conduction band shape for the  sample \#3181.}
\end{indented}
\end{figure}

Both of the structures are symmetric with respect to the middle of central barriers this means that quantum wells are identical for both of them and with great probability their 2DEG densities are matched. The parameters of the structures of both types are listed in Table 1.

\section{Experimental results}

The magneto-transport measurements were performed by means of superconducting magnet, which gives the possibility to get the magnetic fields up to 11 T. The sample was mounted in the anti-cryostat which enables to change the temperature ranging from 0.4 K to 300 K\cite{25}. The external gate voltage was not applied because this field destroys the symmetry of the DQWs profiles. The values of Hall ($U_{xy} (B)$)and transversal $U_{xx} (B)$ voltage were recorded for the two opposite directions of magnetic and electric fields. Therefore, eight records (for magnetic field once going up and then going down) were made and averaged at each single measurement for the transverse magnetoresistance $R_{xx} (B)$ and Hall resistance $R_{xy} (B)$ at given temperature. These curves are shown in Figs. 4a-b and Figs. 5a-b. The pronounced SdH oscillations, which manifest themselves in the $R_{xx} $- magnetic field dependence, as well as Integer Quantum (IQHE) Hall effect ($R_{xy} (B)$-curves) are clearly seen. The SdH oscillations undergo also the beating effect which was observed in our experiments for both types of the structures. We observed four beating-effect nodes for the structure \#2506I (see Fig.4a) and three nodes for the sample \#3181 (Fig. 5a). It is important to mention that no beating-effect was observed in case of the structure with single QW based on the same heterostructures In$_{0.65}$Ga$_{0.35}$As/In$_{0.52}$Al$_{0.48}$As\cite{24}.

\begin{figure}[ht]
\begin{indented}
\item[]
\includegraphics[scale=0.36] {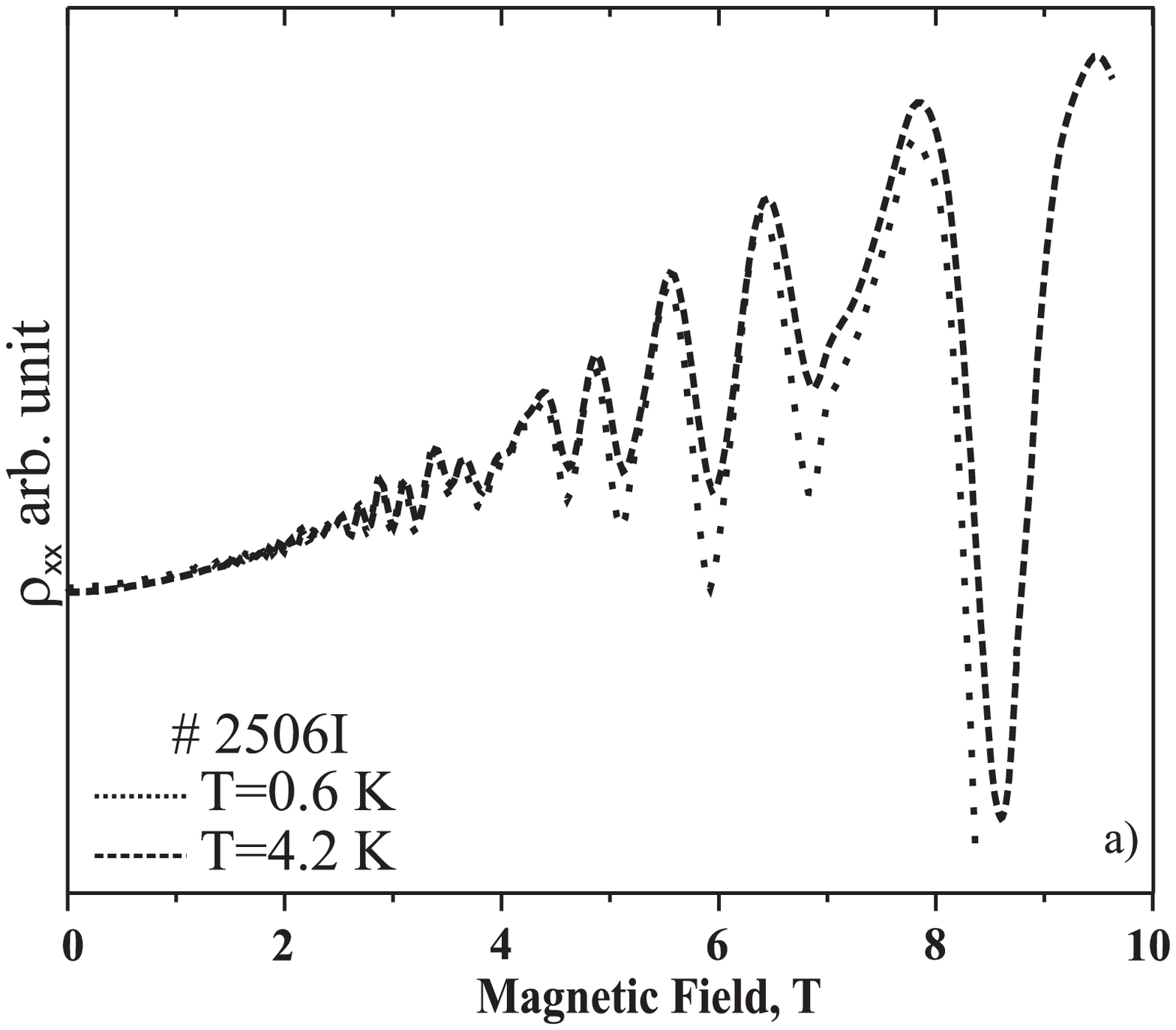}
\includegraphics[scale=0.36] {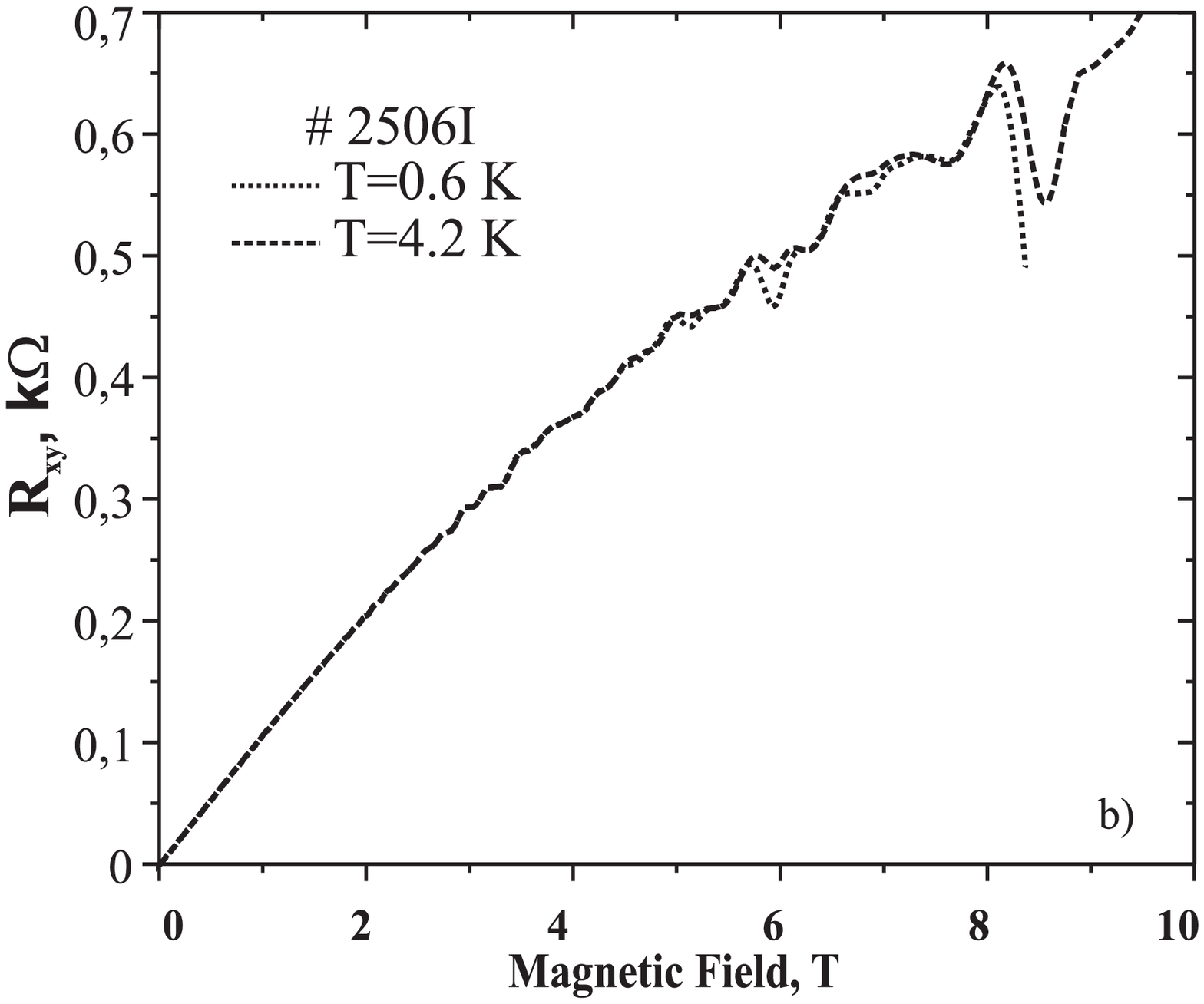}
\caption{\label{fig:epsart}a). The SdH curves for the structure
\#2506I at two temperatures: 0.6 K and 4.2 K. b) The
results of QHE measurements for the structure \#2506I at the
temperatures 0.6 K and 4.2 K.}
\end{indented}
\end{figure}
\begin{figure}[ht]
\begin{indented}
\item[]
\includegraphics[scale=0.36] {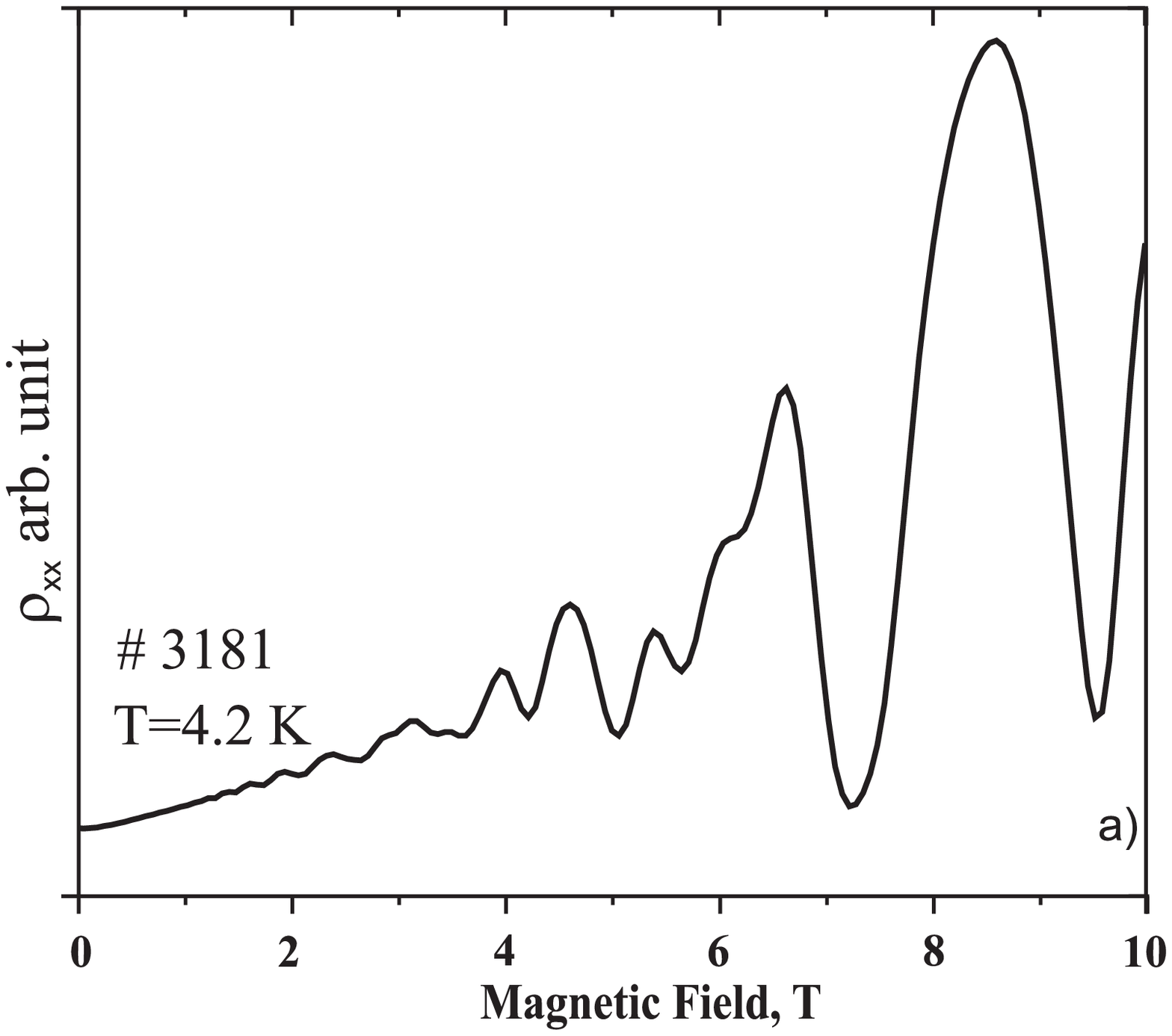}
\includegraphics[scale=0.36] {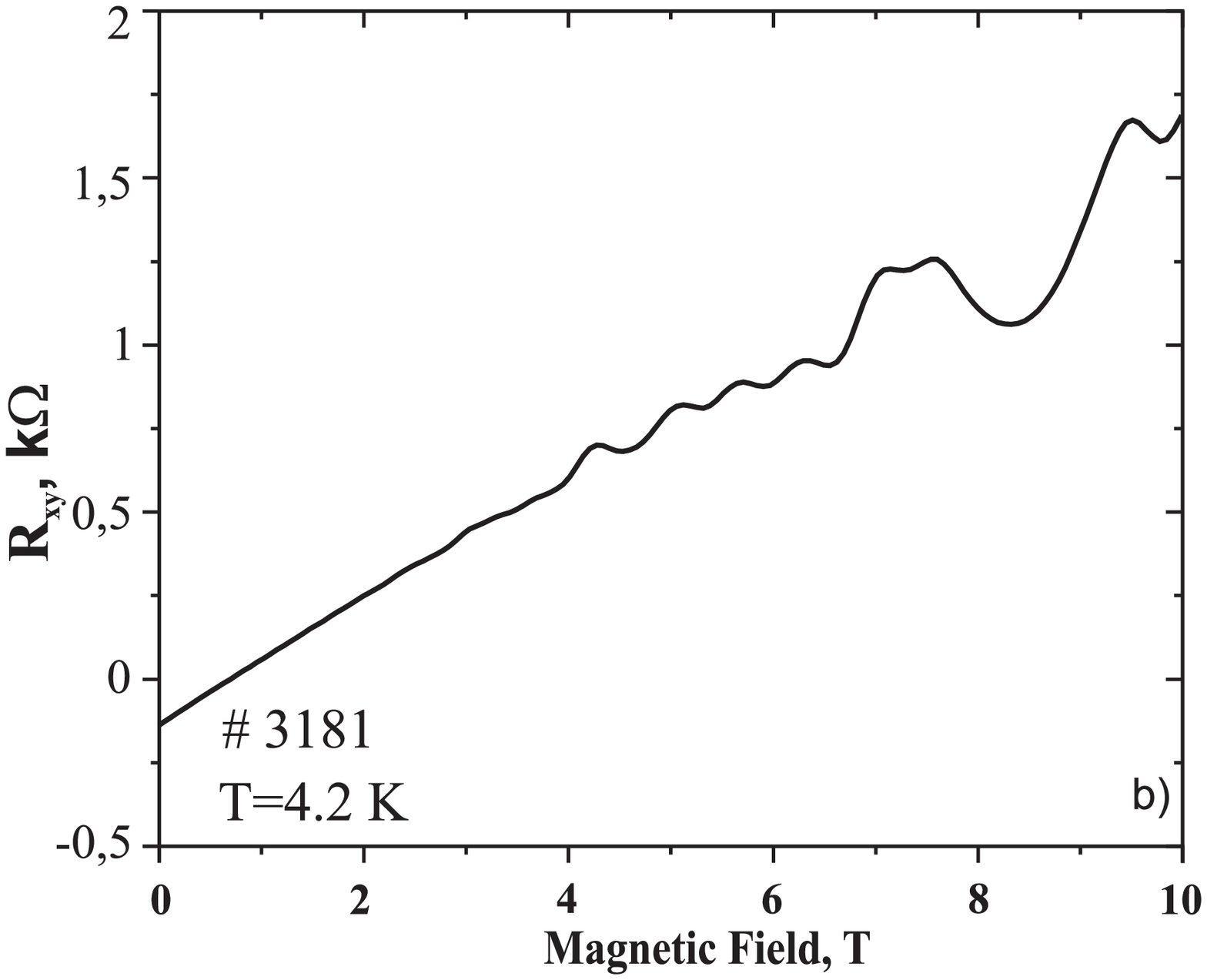}
\caption{\label{fig:epsart}a) The SdH curves for structure \#3181
at 4.2 K. b) The results of QHE
measurements for structure \#3181 at 4.2 K.}
\end{indented}
\end{figure}

 The several plateaux are observed in $R_{xy} (B)$ curves for the magnetic fields up to 6T for the sample \#2506I, as well as for \#3181. It can be seen, that the plateaux observed in the magnetic fields greater than 6 T, are distorted. There could be two possible explanations of this distortion. The first one accounts for the presence of the additional conducting channels, for instance, due to the $\delta $- layers. However, the effect of the additional conducting channels should be more pronounced at 4.2 K, rather than at 0.6 K, while as it is easily seen, the minima of the distorted plateaux are deeper at 0.6 K than those at 4.2 K (see Fig. 4b). The second explanation of the plateau distortion refers to the QWs coupling. The additional maximum on the plateaux was already observed earlier (see Ref. 7).

In Fig. 6 the results of another series of measurements of QHE and SdH oscillations are presented. These results were obtained also for the structure \#2506, but the sample however was cut out from another piece of wafer (see Fig. 2b), that is why we denote it as \#2506II. It is clearly seen that in this case the QHE-plateaux are flat and the Hall resistance $R_{xy} $is quantized at $\hbar /e^{2} $ where the occupation factor is $\nu =(N_{0} +N_{1} )sD$. Here $N_{0} $ is the uppermost Landau level occupied by electrons in the lowest subband (i = 0), $N_{1} $ means the same for the next subband (i = 1), s = 2 is the spin degeneracy, D = 2 is the symmetric degeneracy (symmetric and anti-symmetric states).  At the same time it is necessary to note that together with regularly flat plateaux in QHE for the magnetic field ranging from 3 T up to 8 T, at the magnetic field 8.5 T the small distortion of the plateau is observed. We do not observe the beating effect in SdH oscillations for the sample \#2506II (see Fig. 6). In the next sections we will prove that the beating effect in SdH oscillations is caused by SAS-gap and the difference between two samples of structure \#2506 is related to the difference in carrier density.

\begin{figure}[ht]
\begin{indented}
\item[]
\includegraphics[scale=0.45]{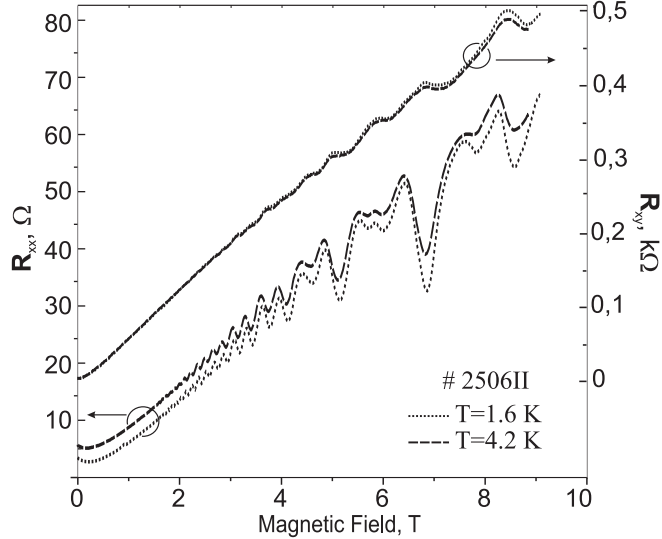}
\caption{\label{fig:epsart} The SdH oscillations and QHE for the second sample \#2506II of the structure \#2056.}
\end{indented}
\end{figure}

The barrier width in the structure \#2506 is equal to 20 nm and the initial, that is in zero-magnetic field, SAS-gap is so small, that can be completely screened by 2DEG (see Ref. 10 and Sect. 5 of this paper) in case of the sample \#2506II. In case of the sample \#2506I the screening is insufficient to cause the collapse of SAS-gap. It is even more valid for the sample \#3181, where the barrier width is 5 nm and SAS-gap is significantly larger. That means that in our case the initial SAS-gap for the two samples of the same \#2506-structure can fluctuate because the electron density slightly changes along the radius of the wafer. Indeed, the electron densities determined by the slop of $R_{xy} (B)$-curve in a small magnetic field for the \#2506II sample is greater then for the \#2506I. It is equal to 4.2$\times 10^{12}$ cm$^{-2}$ for the sample \#2506II and 3.2$\times 10^{12}$ cm$^{-2}$ for \#2506I. As it is shown in Fig. 2c, the distance between two contacts for $V_{xx} (B)$ is about 0,8 mm, while the distance between two samples: \#2506I and \#2506II measured along the radius of wafer is 25 mm. As it follows from our experimental data, the difference in carrier concentrations for two samples \#2506I and \#2506II is about 30\%. Since the distance between the tips of probes measured along the radius of wafer is about 0.8 mm, the carrier concentration change within each of the samples is less than 1\%. Hence, the homogeneity of carrier density is quite satisfactory within the samples used in our experiments. So, for the sample \#2506I we observe at the same time the beating effect and distortion of the plateaux in QHE, while for the sample \#2506II the beating effect is absent and the plateaux are flat within the experimental error for the magnetic field $<8T$.

\section{Simulation of beating effect}

 We carried out the Fourier analysis of the SdH oscillations for both structures, \#2506I and \#2506II, as well as for \#3181. It is clearly seen that in case of sample \#2506I (Fig. 7a) there is predominating main harmonic which is splitted. The similar situation is observed for the sample \#3181 (Fig. 7c) but in this case the magnitude of splitting is larger. That corresponds to beatting effect observed for these two samples. For the sample \#2506II (Fig. 7b) there is one strong harmonic too but it is not so dominate as in previous two cases. It is possible to assume that observed number of harmonics suggests that no less than two subband are occupated by electrons in case of the sample \#2506. 

\begin{figure}[ht]
\begin{indented}
\item[]
\includegraphics[scale=0.60] {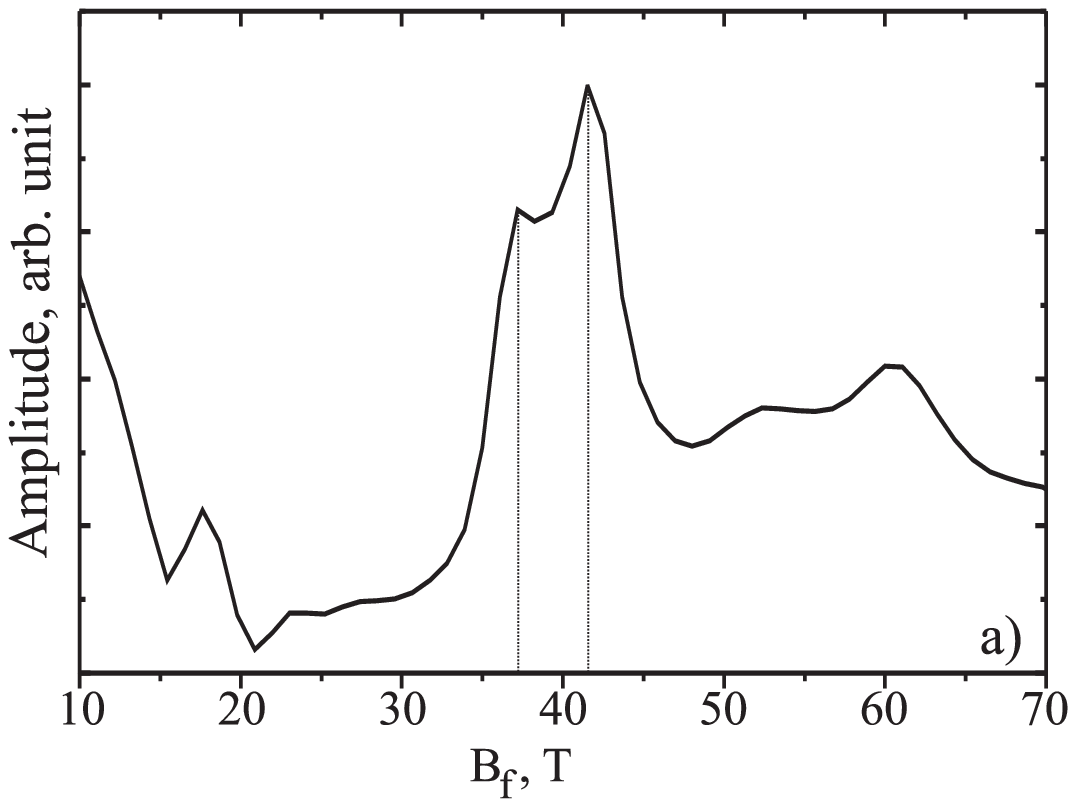}
\includegraphics[scale=0.60] {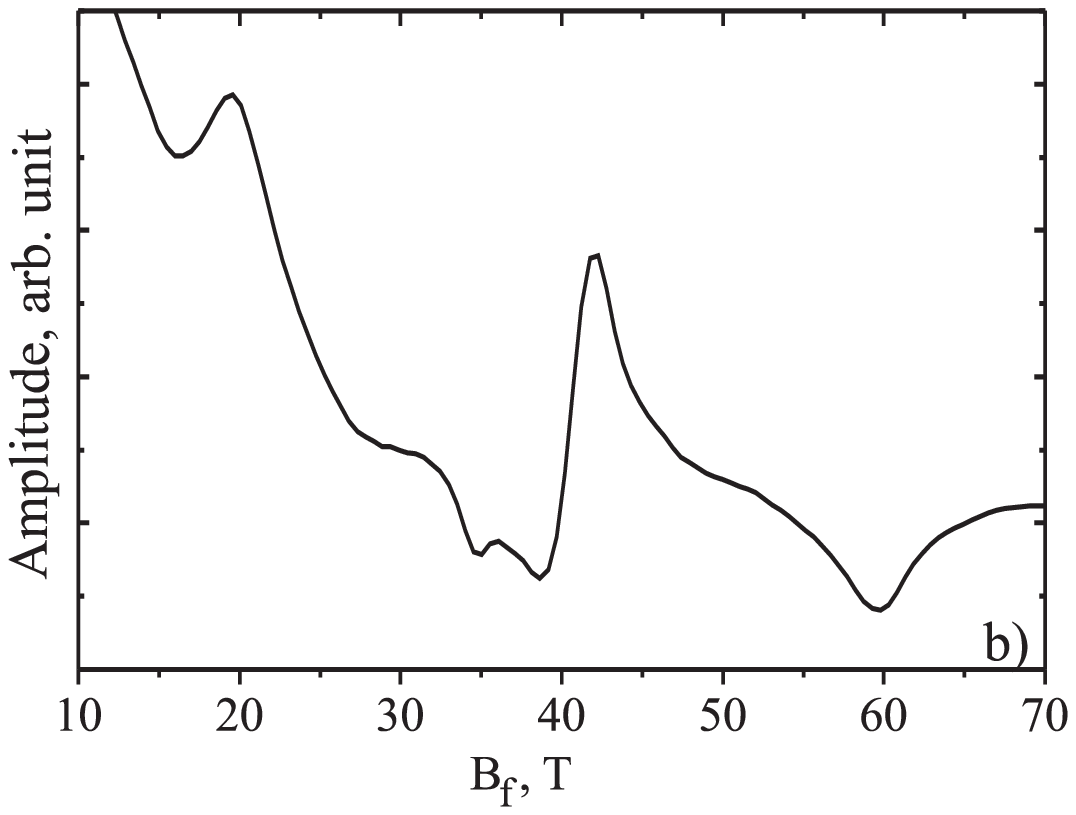}
\includegraphics[scale=0.60] {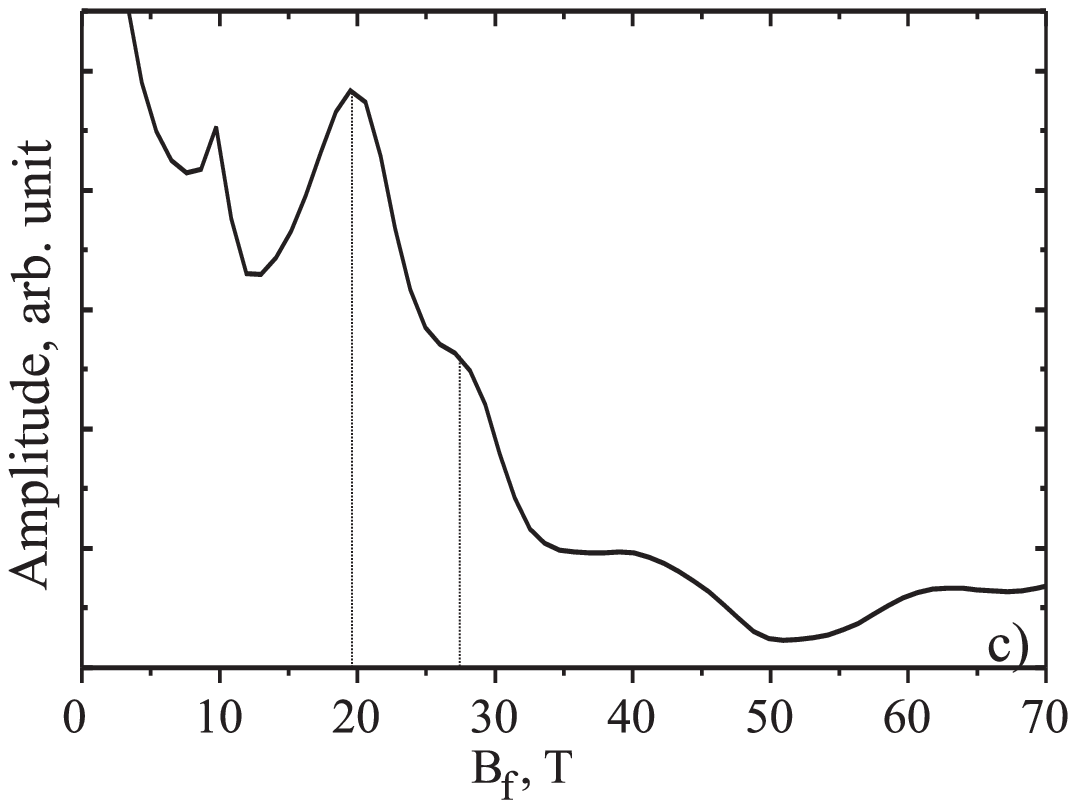}
\caption{\label{fig:epsart} The SdH-oscillations' Fourier transform plotted versus $B_f=[\Delta
(1/B)^{-1}]$: a) for the \#2506I structure, b) for the \#2506II structure and c) for the \#3181 structure.}
\end{indented}
\end{figure}

 From the other hand, as it was mentioned already, our earlier experiments with the structures containing a single quantum well\cite{24}, did not show beating effect which could be attributed to spin splitting. Therefore we can conclude, that another kind of electron states splitting at zero magnetic field also exists. In order to reveal this most important effect, we have used the next method of the experimental SdH curve simulation. The starting point for further analysis is the next expression, which is well-known in the theory of SdH\cite{26}:

\begin{eqnarray}
\frac{\Delta\rho}{\rho_{0}}&=&\sum_{l=1}^{\infty}\frac{5}{2}\biggl(\frac{lP}{2B}\biggr)^
{1/2}\frac{\beta Tm^{'}\cos(l\pi\nu)}{\sinh(\beta Tm^{'}/B)}\times
\nonumber\\ & &\exp(-l\beta T_Dm^{'}/B) \cos2\pi(l/PB-1/8-l\gamma).
\end{eqnarray}

 This formula takes into account the unparabolicity of conduction band as well as spin splitting. Here $\Delta \rho $is the deviation of $\rho $ from background resistivity, $\rho _{0} $ is zero-field resistivity, B is the transverse magnetic field, T is the temperature,  $\beta =2\pi ^{2} k_{B} m_{0} /\hbar e$, $P=\hbar e/E_{f} m^{*} $ stands for the SdH period and, $T_{D} $ is the Dingle temperature.

 The simulation procedure was based on adding only two Fourier harmonics of nearly equal frequency. According to that was mentioned above, the formula (1) can be simplified and reduced to two entries of the Fourier series only. As a result we have:
\begin{eqnarray}
\Delta\rho_{xx}^{1}=\exp\biggl(-\frac{\gamma_{1}}{B}\biggr)\cos
\biggl[2\times(\omega_{1}-kB)\times\frac{\pi}{B}\biggl]
\end{eqnarray}
and
\begin{eqnarray}
\Delta\rho_{xx}^{1}=\exp\biggl(-\frac{\gamma_{2}}{B}\biggr)\cos
\biggl[2\times(\omega_{2}+kB)\times\frac{\pi}{B}\biggl],
\end{eqnarray} 

 where: $\gamma_{1}$ ,$\gamma _{2} $ are the coefficients which are equivalent to $R\beta T_{D} m'$ in Eq. (1), $\omega _{1}$, $\omega _{2} $ are the cyclotron frequencies for two sub-systems of Landau levels. The $kB$-term is included into the argument of cosine in Eqs. (2), (3) in order to have an additional degree of freedom to get better agreement between experimental and calculated curves.

\begin{figure}[ht]
\begin{indented}
\item[]
\includegraphics[scale=0.40] {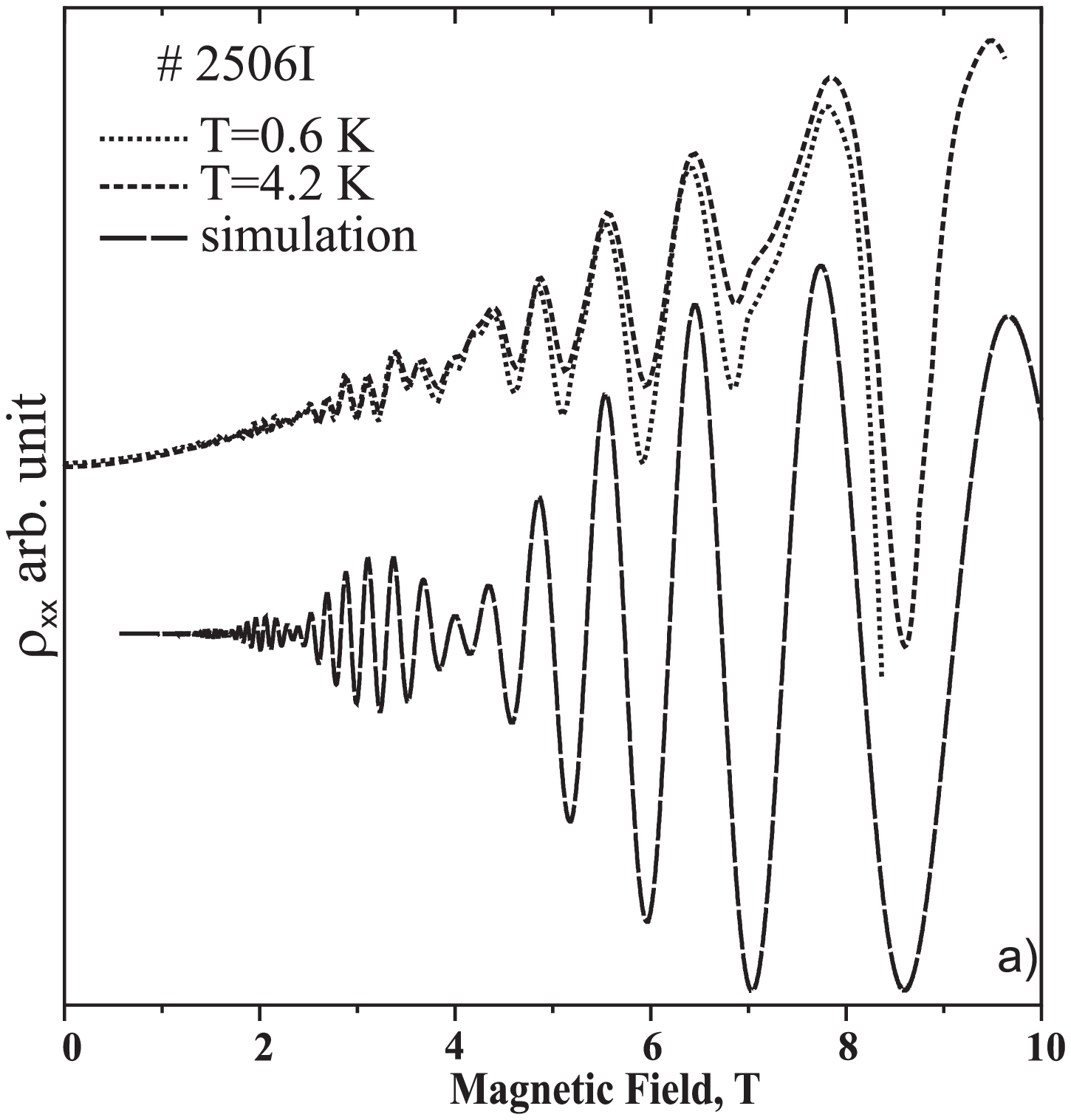}
\includegraphics[scale=0.40] {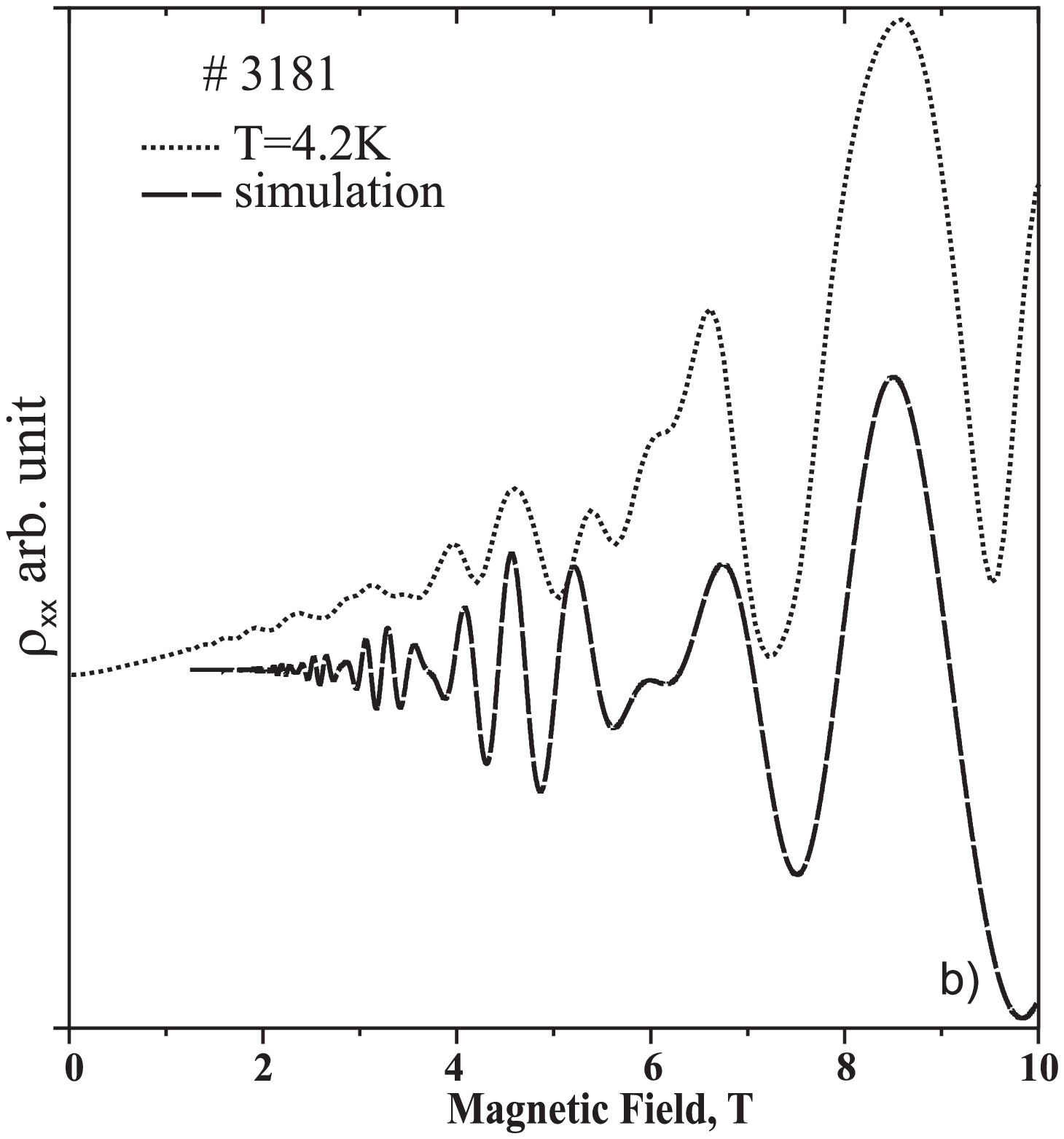}
\caption{\label{fig:epsart}a) The upper curve represents the
experimental data for SdH oscillations (\#2506I). The lower one
represents the results of simulations performed according to the
method described in Sect.4. b) The same is for the structure \#3181.}
\end{indented}
\end{figure}

It turns out, that in order to get better agreement between the experimental results and numerical calculations, one should reduce the oscillation frequencies when the magnetic field increases. These one could confirm the experimental fact that SAS-gap increases if magnetic field increases and if one suppose that zero magnetic SAS-gap causes the beating-effect.

The results of the simulation are shown in Figs. 8a and 8b for
the samples \#2506I and \#3181, respectively. The fitting parameters
used in the simulations, were chosen to be: $\gamma_{1}=8.8$,
$\gamma_{2}=8$, $\omega_{1}=34.9$, $\omega_{2}=40.27$, $k=0.095$, and
$\gamma_{1}=9$, $\gamma_{1}=8.8$, $\omega_{1}=30.8$,
$\omega_{2}=23.8$, $k=0.095$ for these two samples, respectively. In
this way we have obtained two oscillation components for each
SdH-curve. These curves are shown in Figs. 9a, 9b and 10a, 10b. They should be interpret using of Landau levels energy
calculates carried out for the symmetric and anti-symmetric states separately,
by means of the best fit procedure.

\section{Calculation of Landau-level energy}

 In our calculations we have used the model proposed in Ref. 10, where the symmetric DQWs separated by the middle barrier were considered. An external magnetic field was applied perpendicular to the planes of QWs. The electron's behavior was described by the Schr\"odinger equation which included the DQW potential, as well as the self consistent Hartree potentials. The in-plane movement in each of the subbands of the DQW was supposed to be Landau quantized. The total electron energy which includes the contributions of the in-plane as well as vertical electron motion, can be written as:
\begin{equation}
E_{ni}=\biggl(n+\frac{1}{2}\biggl)\hbar\omega_{c}+E_{i}+V_{ni}^{F},
\end{equation} 
where $V_{ni}^{F} $ is the exchange energy (which involves screening) given by Eqs. (12)-(18) of Ref. [10], for DQW based on the GaAs well-material and Al$_{0.3}$Ga$_{0.7}$As barrier-material with the well width of 14 nm and the barrier width of 3 nm. The authors of Ref. 10 have found an approximately linear B-dependence of the tunneling gap for the magnetic fields $0<B<9T$. One can conclude therefore, that the total splitting of energy states of two QW's $\Delta _{t} $, consists of two parts: constant $\Delta _{SAS} $, determined by the overlapping of the electron wave functions of two QW's, and changeable tunneling term $V_{ni}^{F} $, involving screening factor:

\begin{equation}
\Delta_{t}=\Delta_{SAS}+V_{ni}^{F}.
\end{equation} 

The last term, being strongly reduced when B increases up to the values for which the quantum limit occurs, can be written as:

\begin{equation}
V_{ni}^{F}=-(K_{0}-kB),
\end{equation}

where $K_{0} $is the maximum screening factor in zero magnetic field depending mainly on the electron densities and $-(kB)$is the changeable screening factor decreasing with $B$ up to the value when the first Landau level in two QWs becomes to be occupied. As it was observed experimentally\cite{10}, the total energy gap $\Delta _{t} $between symmetric and anti-symmetric states in DQW is proportional to the magnetic field. 

 We adapted this model for our particular case of symmetric In$_{0.65}$Ga$_{0.35}$As/In$_{0.52}$Al$_{0.48}$As DQW structure (\#2506I) assuming rectangularity of QWs. The last one enables to calculate the Landau level energies according to the formula\cite{27}:

\begin{equation}
\frac{(E^{'}-E_{\bot})(E_{g}+E^{'}+E_{\bot})}{E_{g}}=\frac{\hbar^{2}\pi^{2}(i+1)^{2}}{2m_{c}^{*}a^{2}r},
\end{equation}
\begin{equation}
E_{\bot}=-\frac{E_{g}}{2}-\frac{E_{g}}{2}\times \sqrt{1+\frac{4\mu_{B}B}{E_{g}}\biggl[f_{1}
\frac{m_{0}}{m_{c}^{*}}\left(n+\frac{1}{2}\right)\pm\frac{1}{2}g_{0}^{*}f_{2}}\biggr],
\end{equation}
where
\begin{eqnarray}
&&f_{1}=\frac{E_{g}+\frac{2}{3}\Delta}{E_{\bot}+E_{g}+\frac{2}{3}\Delta} \nonumber\\
&&f_{2}=\frac{(E_{g}+\Delta)(E_{\bot}+E_{g}+\frac{2}{3}\Delta)}{(E_{g}+\frac{2}{3}\Delta)(E_{\bot}+E_{g}+\Delta)}
\end{eqnarray}
where $E^{'}=E\pm(\Delta_{SAS}^{s}+k^{'}E_{\bot})$, $E_{g}$ is the energy gap (between the top of valence band and the edge of conduction band) and  is spin-orbital splitting, $m_{0} $ is the electron mass in a vacuum, $m_{c}^{*} $ is the effective electron mass corresponding to the edge of the conduction band, $\mu _{B} $ is the Bohr magneton, $n=0,1,2 ...$ stands for Landau levels, $i=0,1,2 ...$ is the number of sub-bands, $E_{\bot } $ is the Landau level energy for bulk semiconductor (we use here the W. Zawadzki's notation\cite{27}) and $E$ is the unknown $n^{th}$ Landau level's energy for the  $i^{th}$ sub-band in QW. The parameter $\Delta _{SAS}^{s} $ is the energy gap determined by the overlapping of the wave functions, which is diminished by the screening at zero magnetic field as it follows  from (5) and (6):

\begin{equation}
\Delta_{t}=(\Delta_{SAS}-K_{0})+k'B\equiv\Delta_{SAS}^{s}+k'B.
\end{equation}

Thus, the SAS-gap is proportional to the magnetic field B. The coefficient of proportionality $k'$ can be considered as fitting parameter and is equal to $2k=0.19$, as it was determined in previous Section. We assume the triangular potential shape in case of asymmetric DQW (\#3181) with one donor $\delta$-layer in the barrier on one side of each QW. It means that Eq. (4) has to be modified for QW with the triangular potential (see Ref. 27):

\begin{equation}
(a+b)a^{1/2}b^{1/2}+(a-b)^{2}\ln\left|\frac{b^{1/2}-a^{1/2}}{(b-a)^{1/2}}\right|=
\biggl[\frac{E_{g}^{*}}{2m_{c}^{*}}\biggr]^{1/2}4eF\hbar \pi(i+3/4),
\end{equation}

where: $a=E^{'}-E_{\bot}$; $b=E_{g}^{*}+E^{'}+E_{\bot}$; $E^{'}=E\pm(\Delta_{SAS}^{s}+k^{'}E_{\bot})$; $F$ is the electric field caused by the existence of the interface and which is determined by the potential $U=eFz$. The magnitude of $E_{\bot } $ is determined in the same way as previously, by means of Eqs. (8) and (9). The equations (7) and (11), together with (8) and (9) allow to calculate the Landau levels (LL's) energies in rectangular and triangular DQW with the spin-splitting and tunneling gaps variable in magnetic field taken into account. We used then the results of these calculations to account for the SdH oscillations and to find the parameters of 2DEG for both of the structures (the Fermi level (FL), carriers density etc.). 

\begin{table}[h]
\caption{\label{label}Band-structure parameters used for the LL's energies
calculation, where $E_{g}$-energy gap, $m_{c}^{*}/m_{0}$-the ratio of electron mass and effective
electron mass, $\Delta$-spin-orbital splitting.}
\begin{indented}
\item[]
\begin{tabular}{@{}lllll}
\br
$x$     &$E_{g}, [eV]$      &$m_{c}^{*}/m_{0}$& $g_{c}^{*}$     & $\Delta, [eV]$\\
\mr
0.65    & 0.723             & 0.0229           & -9.90          & 0.355 \\
\br
\end{tabular}
\end{indented}
\end{table}

The calculations of the LL's for the sample \#2506I were performed using the band parameters, determined previously for the sample with the single quasi-rectangular QW[24]. These parameters are listed in Table 2, they are the same for both of the samples. As for the value of $\Delta _{SAS}^{s} =1$ meV for the \#2506I structure and $\Delta _{SAS}^{s} =5.5$ meV for the \#3181 structure.

\section{Discussion}

\subsection{The role of SAS-gap.}

 As was mention about since no beating-effect in SdH oscillations was observed in case of a structure with a single QW\cite{24}, we presume that this effect cannot be attributed to the spin-splitting of Landau levels. There are two alternative explanations of this effect. The first one is due to the existence of the SAS-gap at zero magnetic field. The second one is due to different carrier densities in two QWs (so called mismatching of carrier densities). It is well known that for different carrier densities in two QWs, the Fermi levels in both of them are also different if the tunneling effect is negligible. As a results of this difference, the beating effect could be observed. It is worthy to mention however, that according to Ref. 6 where the mismatching of electron densities were created in controllable way, the beating effect was not observed. We believe that in our  case the beating effect is not the result of difference in carrier densities in two QWs, but is the result of the existence of SAS-gap in zero magnetic field. There are three evidences for that. The first one is the symmetry of the structures in questions with respect to the center of the middle barrier: both of the QWs were produced in a way to make them identical. The second one is that the width of the central barrier is small enough in order to equalize the carrier densities in both of the QWs of the structure \#2506, as well as \#3181. Note that in case of \#3181-structure the barrier width is equal to 5 nm and in case of \#2506-structure is 20 nm. If the beating effect were caused by the difference in electron densities, it would be more pronounced in case of \#2506- structure then in \#3181 and the period of beating (versus $1/B$) would be smaller for the structure \#2506, while in our experiment it was just the opposite. And at last the third evidence. In two samples of the same structure (\#2506I and \#2506II) the electron densities are different, but the difference is comparatively small. If the beating effect were caused by the difference in electron densities in two QWs, it would be impossible to explain, why the beating effect disappears when the electron densities increase a bit in one of the samples (\#2506II). And the contrary, the disappearing of beating in \#2506II can be very naturally explained, if we attribute it to the existence of SAS-gap in zero magnetic field, because the small increase in electron density leads to the collapse of small SAS-gap due to screening. 

Therefore, the non-zero SAS-gap causes the SdH- oscillations at two different frequencies (versus $1/B$) and leads to beating effect.

In the next sub-section we will show that our experimental data related to beating effect can be sufficiently well interpreted, if one takes into account the accurately calculated LL energies and introduces two quasi-Fermi levels characterizing two electron sub-systems which belong to symmetric and anti-symmetric states in DQW-structure.

\subsection{Two quasi-Fermi levels.}

 The calculations of LL's energies were performed according Eq. (7) for the sample \#2506I and Eq. (11) for the sample \#3181 using parameters shown in table 2. The results of our calculations for the sample \#2506I are shown in Fig. 9a (for the symmetric states) and in Fig. 9b for the anti-symmetric ones. The calculated energies are plotted versus magnetic field for two subbands (i$=$1,2). Usually the Fermi level (FL) position are determined by LL's and FL crossings corresponding to the maxima of two components of SdH oscillations, acquired by simulation of beating effect. It is worth mentioning that, as it is seen in the Figs. 9a, 9b, the FL-positions are perfectly well fitted to the oscillation curves and are unique, since they are correspond to the regularly spaced crossings of LL's, equidistant with respect to the inverse magnetic field.

\begin{figure}[ht]
\begin{indented}
\item[]
\includegraphics[scale=0.43] {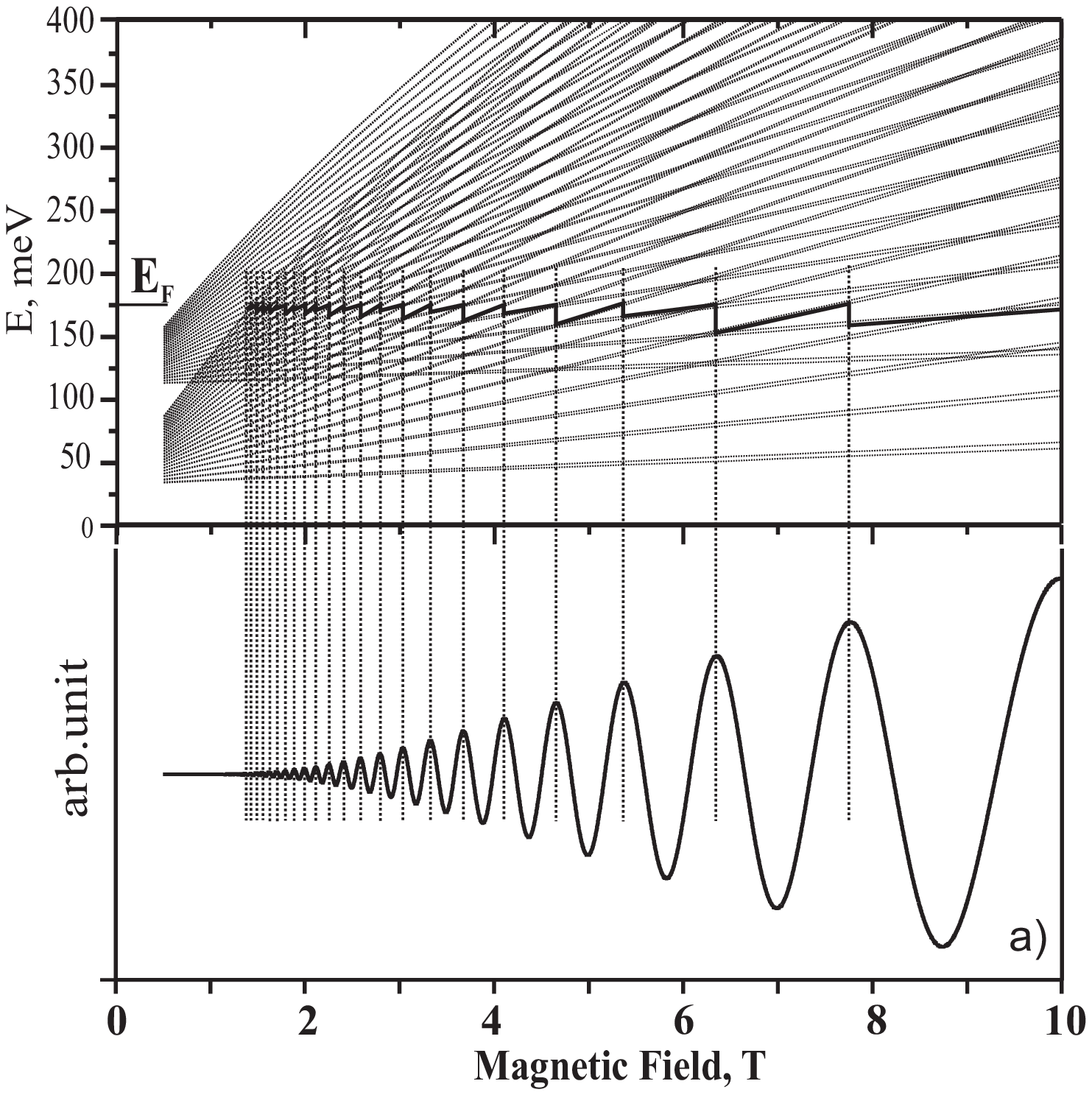}
\includegraphics[scale=0.43] {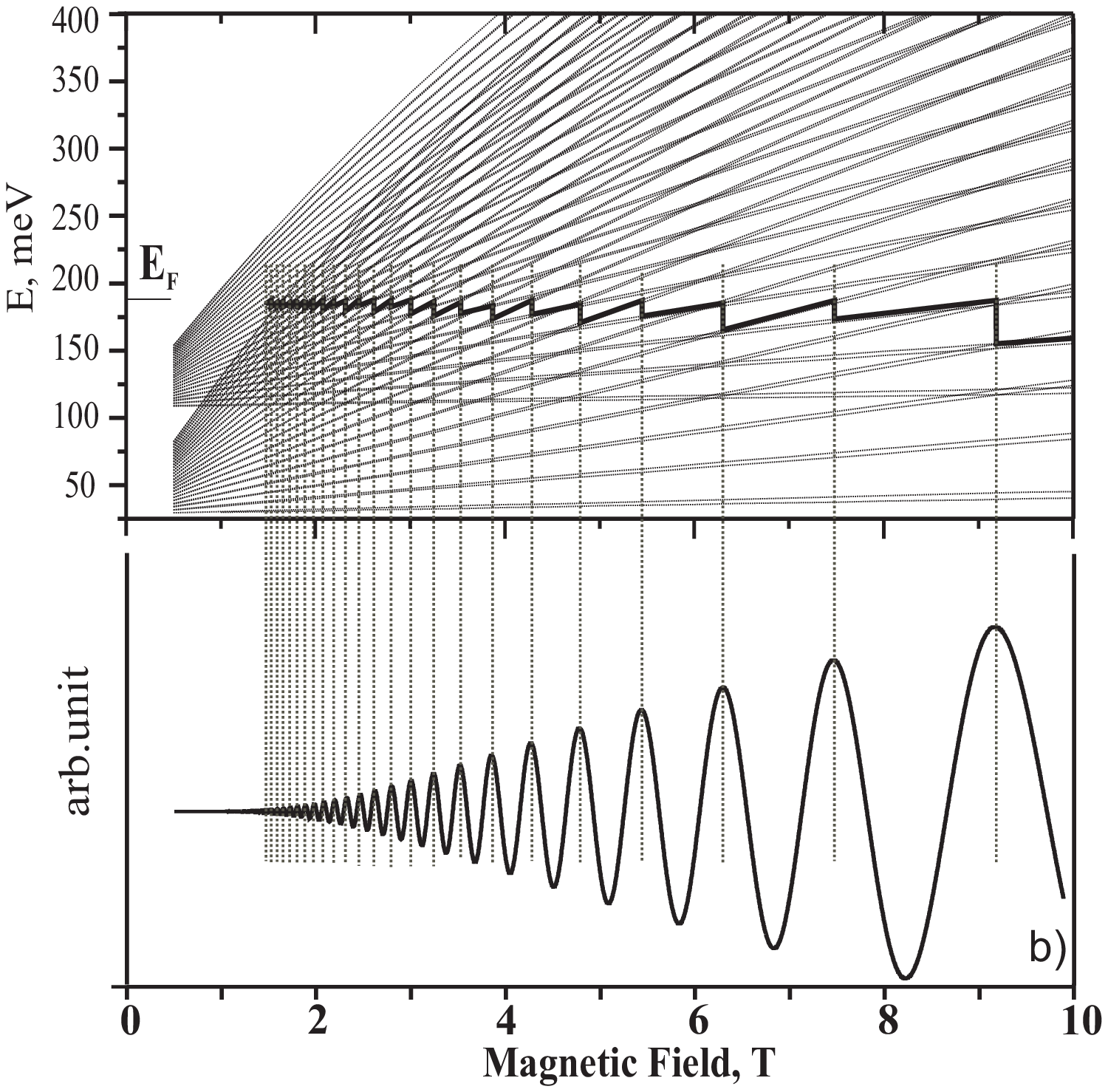}
\caption{\label{fig:epsart} These figures represent the best fit of
the Fermi level energy corresponding to: a) symmetric part of the
SdH oscillations b) anti-symmetric part of the SdH oscillations, in
case of the structure \#2506I.}
\end{indented}
\end{figure}
\begin{figure}
\begin{indented}
\item[]
\includegraphics[scale=0.44] {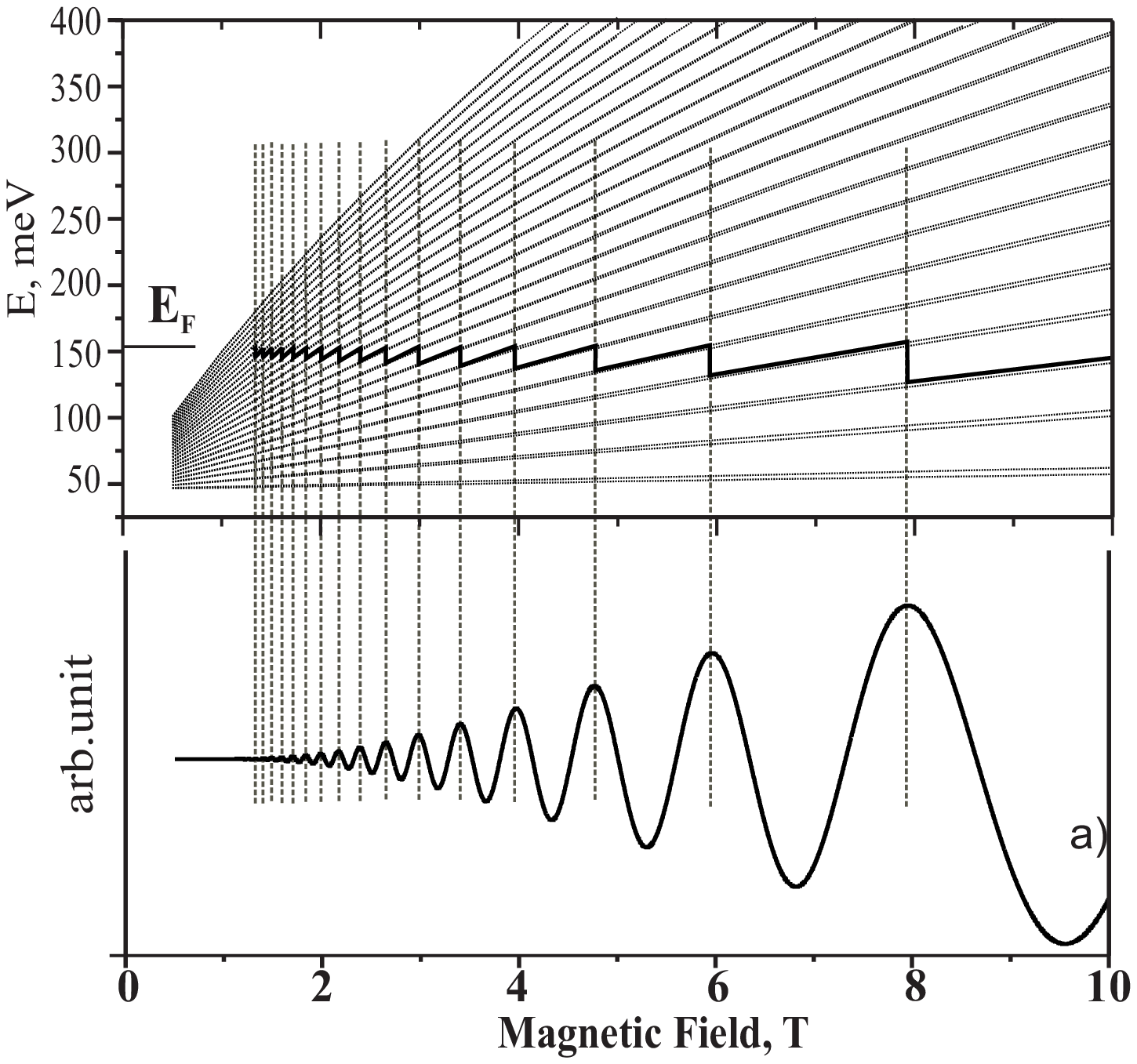}
\includegraphics[scale=0.44] {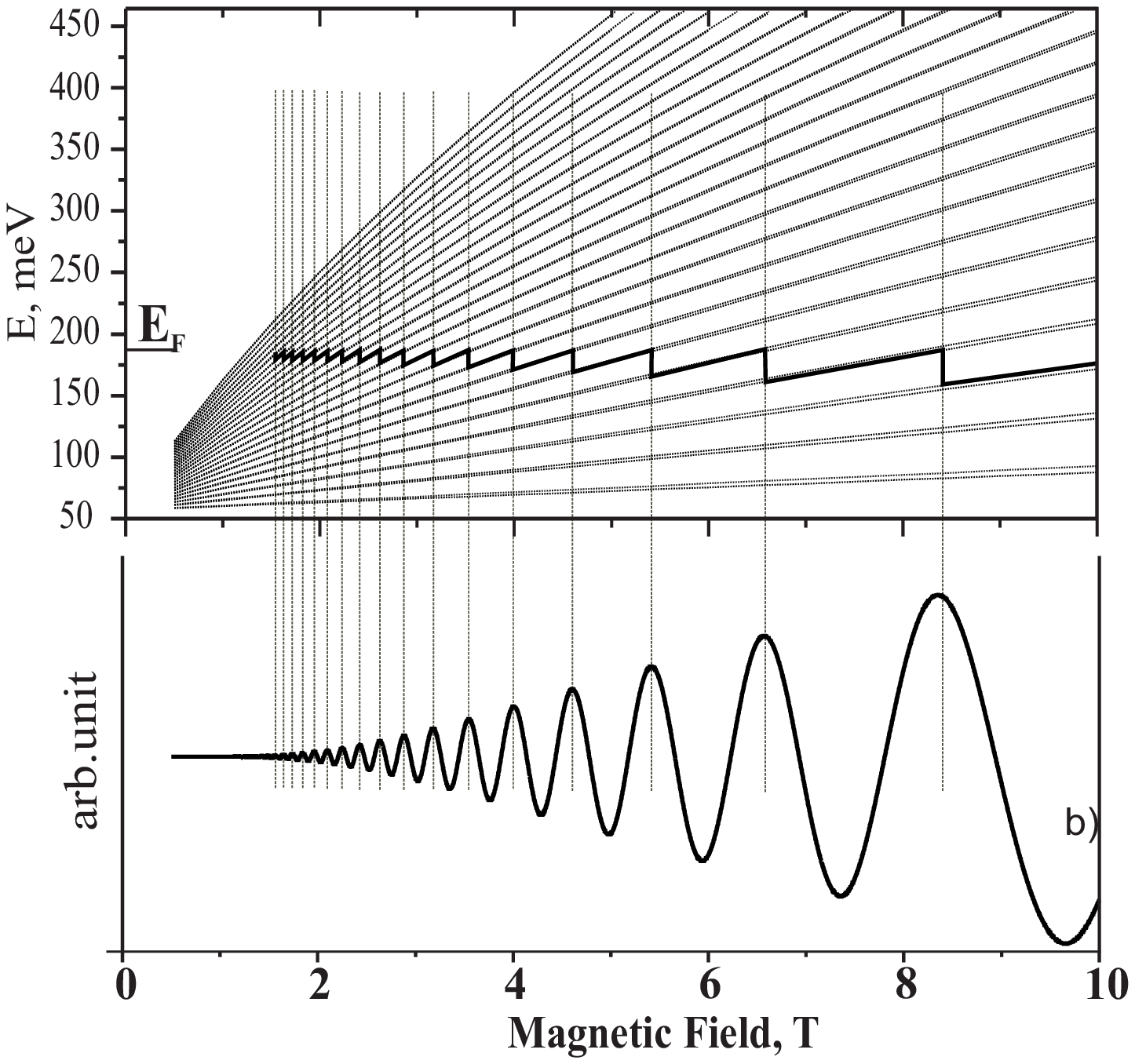}
\caption{\label{fig:epsart} These figures represent the best fit of
the Fermi level energy corresponding to: a) symmetric part of the
SdH oscillations b) anti-symmetric part of the SdH oscillations, in
case of the structure \#3181.}
\end{indented}
\end{figure}

 The FL displacement up or down, destroys this regularity immediately, due to the superposition of the two space-quantized subbands. We should admit that we could not achieve the satisfactory agreement between our experimental data and the theoretical predictions about the SdH oscillation peaks positions, if we tried to use the single Fermi level common for both types of the states, symmetric and anti-symmetric ones. As it is clearly seen in Fig. 9a and 9b, the Fermi levels for two oscillation components are different, and the difference is equal to 12 meV. The similar is true for the sample \#3181. In case of this sample only one subband is occupied, therefore to determine FL is simpler problem to solve. In spite of this, it was impossible to find a single Fermi level which would be common for both sub-systems, symmetric and anti-symmetric. It turns out that the difference between two Fermi levels for the symmetric and anti-symmetric states in this case is equal to 32 meV (see Fig. 10a and 10b, respectively).

 It is interesting to note, that in case of the sample \#2506I, the value of total splitting $\Delta _{t} $ is approximately equal to 15 meV for n$=$0, i$=$0, in a magnetic field of about B$=$10 T, and should increase if the field increases up to about 30 T, where the predicted quantum limit should occur. The obtained values of the FL energies give us the possibility to estimate the density of carriers for both structures. These values are 3.8$\times 10^{12}$ cm$^{-2}$ for the sample \#2506I (good agreement with date obtained by the slop of $R_{xy} (B)$-curve in small magnetic field) and 2.5$\times 10^{12}$ cm$^{-2}$ for the sample \#3181. In the first case this value is a bit greater than the technologists set it up to be, and in the second case is a bit smaller. So, our experimental data force us to introduce the concept of two quasi-Fermi levels, by means of which two electron sub-systems can be characterized. These two sub-systems are the electrons belonging to two different types of states, symmetric and anti-symmetric ones. One can encounter the concept of quasi-Fermi levels in different contexts for instance, in semiconductor physics. The quasi-Fermi levels for electrons and holes very often are used; quasi-Fermi levels for the electron states with $k_{+}$ and $k_{-} $($k$are for the electron wave vectors along say, $+x$ and $-x$ axes) are introduced to explain the peculiarities of IQHE\cite{28,29}, and so on. Generally speaking this concept simply means, that one can treat two sub-systems as weakly interacting and having their own rate of establishing the equilibrium state. Therefore, our next step is to argue that indeed, we could interpret our data, introducing the two quasi-Fermi levels which characterize two separate electron sub-systems.

\subsection{Electron-electron interaction for the two sub-systems: symmetric and anti-symmetric states.}

Suppose now that the 2DEG in the DQW structure under consideration is off the thermodynamic equilibrium state and the deviations from the equilibrium are small enough. Obviously, these deviations are caused by the current which flows through the structure. Then the electron distribution functions, one of which corresponds to the symmetric state $f^{(a)}$ and the other to anti-symmetric one $f^{(b)} $, can be represented as follows:

\begin{equation}
f^{(a,b)}=f_{0}^{(a,b)}\left(\vec{k}\right)+\delta
f^{(a,b)}\left(\vec{k}\right)
\end{equation}
where
\begin{eqnarray}
&\delta f^{(a)}\left(\vec{k}\right)=\delta_{\vec{k}\vec{k_{0}^{'}}}\delta f^{(a)}\left(\vec{k_{0}}\right),\nonumber\\
&\delta
f^{(b)}\left(\vec{k}\right)=\delta_{\vec{k}\vec{k_{0}^{'}}}\delta
f^{(b)}\left(\vec{k_{0}}\right),
\end{eqnarray}
Here the superscript $a$  stand for the symmetric and $b$ for the anti-symmetric states (see App. A where the wave function for this states are considered). Denote the electrons belonging to sub-system $a$ by means of $\psi _{a}$, and belonging to sub-system  $b$ as $\psi _{b} $ and define the rate of establishing the equilibrium state in each sub-system as: 

\begin{eqnarray}
\frac{1}{\tau^{(a,b)}\left(\vec{k}\right)}=
\sum_{\vec{k^{'}}}\Biggl\{W_{\vec{k}\rightarrow
\vec{k^{'}}}^{(a,b)}\biggl[1-f_{0}^{(a,b)}\left(\vec{k^{'}}\right)\biggr]+f_{0}^{(a,b)}\left(\vec{k^{'}}\right)W_{\vec{k^{'}}\rightarrow\vec{k}}^{(a,b)}\Biggr\}=\nonumber\\
\sum_{\vec{k^{'}}}W_{\vec{k}\rightarrow\vec{k^{'}}}^{(a,b)}\frac{1-f_{0}^{(a,b)}\left(\vec{k^{'}}\right)}{1-f_{0}^{(a,b)}\left(\vec{k}\right)},
\end{eqnarray}

where $W_{\vec{k}\to \vec{k}'\left(\vec{k}'\to \vec{k}\right)}^{(a,b)} $ are the probabilities for the electrons belonging to the sub-system $a$ or $b$ to be scattered from the state $\vec{k}$to the state $\vec{k}'$and vice versa (see Fig. 11). These probabilities are determined by the next formula:

\begin{equation}
W_{i\rightarrow f}^{(a,b)}= \frac{2\pi}{\hbar}\left|\left \langle
\Psi_{f}^{(a,b)} \left|\hat{V}\right|\Psi_{i}^{(a,b)}\right\rangle^{2} \right|
\delta \left(E_{i}^{(a,b)}-E_{f}^{(a,b)}\right),
\end{equation}

where ${\left| \Psi _{i}^{(a,b)}  \right\rangle} $ ,  ${\left| \Psi _{f}^{(a,b)}  \right\rangle} $ are the initial and final states, $E_{i}^{(a,b)} $, $E_{f}^{(a,b)} $ are their energies and $\hat{V}$ is the operator, corresponding to the perturbation responsible for the transition. It is well-known that mainly the electron-electron ($e-e$) collisions lead to establishing the equilibrium states within the electron gas in a semiconductor at low temperatures\cite{30}. It means that whatever the initial distribution function is, the final distribution can be described by displaced Fermi (or displaced Maxwellian) distribution function. However, for establishing of such distribution, an important condition has to be satisfied:  $e-e$ collisions should be more frequent than the electron-phonon ($e-ph$) collisions. In other words,  $\tau _{ee} <<\tau _{e-ph} $, where $\tau _{ee} $ and $\tau _{e-ph} $ are the $e-e$ and $e-ph$ scattering times, respectively. Since the temperatures at which our experiments were carried out, were very low, we can suppose the last condition to be fulfilled. Now in order to calculate $\tau _{ee} =\tau ^{(a,b)} \vec{k}$ for the two sub-systems, we have to calculate $W_{\vec{k}\to \vec{k}'}^{(a,b)} $, the corresponding transition probabilities. Therefore, we should analyze the probabilities for two electrons to transit from the states $\left(l_{1} \vec{k}_{1} ,l_{2} \vec{k}_{2} \right)^{(a,b)} $ to the states $\left(l'_{1} \vec{k}'_{1} ,l'_{2} \vec{k}'_{2} \right)^{(a,b)} $ where $l_{1} l'_{1} ,l_{2} ,l'_{2} $ stand for the quantum numbers, other then $\vec{k}$.

\begin{figure}[ht]
\begin{indented}
\item[]
\includegraphics[scale=0.40]{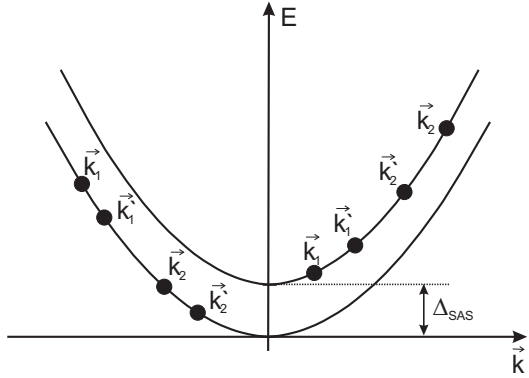}
\caption{\label{fig:epsart}Diagram, representing the
electron-electron scattering in case of two sub-bunds, symmetric and
anti-symmetric.}
\end{indented}
\end{figure}
 
Notice that the subscripts 1,2 stand for the electrons 1 and 2 in a pair, while the superscripts (a,b) are to distinguish symmetric and anti-symmetric states of the electrons in DQW structure. Then 

\begin{equation}
W_{(a,b)\rightarrow \left(1^{'}2^{'}\right)}^{(a,b)}=
\frac{2\pi}{\hbar}\left|M_{(1,2)\rightarrow \left(1^{'},2^{'}\right)}^{(a,b)}\right|^{2}\times
\delta\left(\epsilon_{1}^{(a,b)}+\epsilon_{2}^{(a,b)}-\epsilon_{1}^{'(a,b)}-\epsilon_{2}^{'(a,b)}\right).
\end{equation}

Here we introduced the following notations $l_{1}k_{1}\equiv1$,
$l_{1}^{'}k_{1}^{'}\equiv1^{'}$ etc; $\epsilon_{1}^{(a,b)}$,
$\epsilon_{2}^{(a,b)}$, $\epsilon_{1}^{'(a,b)}$,
$\epsilon_{2}^{'(a,b)}$, are the electron energies in the initial and final states, respectively. It is necessary to calculate the matrix element $M_{(1,2)\to (1',2')}^{(a,b)} $ In Eg. 16 to show the different rates of establishing of the quasi-equilibrium states in two sub-systems. The calculation of two elements $M_{(1,2)\to (1',2')}^{(a)} $ and $M_{(1,2)\to (1',2')}^{(b)} $ is caring out in App. B.

 By inspecting Eqs. (B.3) and (B.4), even without doing out the evaluation of these integrals, one can easily see that $M_{(1,2)\to (1',2')}^{(a)} \ne M_{(1,2)\to (1',2')}^{(b)} $ and hence, the same is true for $\tau _{ee}^{(a,b)} $: $\tau _{ee}^{(a)} \ne \tau _{ee}^{(b)} $. That is, the rates of establishing of the quasi-equilibrium state under electron-electron scattering in two sub-systems, corresponding to the symmetric and anti-symmetric subbands, are different.

 The same arguments are also valid, if we consider the electron-phonon interaction. Moreover in case of electron-phonon interaction the formulae analogous to (B.3) and (B.4) are even simpler, because in this case the initial as well the final electron states are one-particle states, but not the two-particles ones, as in case of electron-electron scattering. Anyway, one can state that both  $\tau _{ee}^{(a,b)} $ and $\tau _{e-ph}^{(a,b)} $ in two sub-systems corresponding to the symmetric and anti-symmetric states,  are different: $\tau _{ee}^{(a)} \ne \tau _{ee}^{(b)} $; $\tau _{e-ph}^{(a)} \ne \tau _{e-ph}^{(b)} $ .

 The only thing which also has to be proven, is that these two sub-systems are weakly interacting. To this end let us consider the transition matrix element $M_{(1,2)\to (1',2')}^{(c)} $ where (1,2) belong to initial symmetric state $\psi _{i}^{a} $ and ($1',2'$) - to final anti-symmetric state $\psi _{f}^{b} $. Then, under the integral sign in (B.3) one should have the product of two functions $\varphi _{a} (z_{1} )$, $\varphi _{b} (z_{1} )$, the one is even and another one is odd. Notice also, that $U^{ee} $is an even function of their variables. Hence, $M_{(1,2)\to (1',2')}^{(c)} $ should be very nearly to zero due to parity selection rule, and it means that the electron-electron scattering does not mix the states of the different parities. Of course, electron-phonon interaction can and usually do them mixed. However, the temperatures at which our experiments were carried out were so low and the electron densities in QW so high, that we the main scattering mechanism is the electron-electron scattering, and that two sub-systems belonging to symmetric and anti-symmetric states are weakly interacting in that sense.

\section{Summary}

 We have studied the parallel magneto-transport in DQW-structures of two different potential shapes (quasi-rectangular and quasi-triangular) and have found an important contribution of the symmetric properties of the charge carriers eigenstates to the magneto-transport phenomena. The beating effect occurred in SdH oscillations was observed for both types of DQW. To explain this effect, we developed a special scheme for the Landau levels energy calculations based essentially on the model proposed in Ref.[10] and carried out the necessary simulations of beating effect. In order to obtain the agreement between our experimental data and the theoretical predictions concerning SdH-oscillations, we introduced two different quasi-Fermi levels, which characterize sub-systems of symmetric and anti-symmetric states in DQWs. The existence of two different quasi Fermi-levels simply means, that in DQW's consisted of two identical InGaAs QWs separated by InAlAs-barriers no wider than 20 nm, the total electron system can be considered as divided into two sub-systems characterized by symmetric and anti-symmetric wave functions respectively. One can treat then these two sub-systems as weakly interacting in the sense that electron-electron scattering does not mix them and that they are described by their own distribution functions characterized by their own quasi-Fermi levels. 

\appendix
\section*{Appendix A}
\setcounter{section}{1}

 The quantum magnetotransport in DQW can be treated in the framework of the  effective mass approximation\cite{29}. The main equation of the effective mass model is of the form:

\begin{equation}
\biggl[E_{s}+\frac{\left(i\hbar\nabla-\vec{eA}\right)^{2}}{2m^{*}}+U(z)\biggr]\psi(\vec{r})=E\psi(\vec{r}),
\end{equation}

where $E_{s} =E_{c} +\varepsilon _{1} $, $E_{c} $ stands for the bottom of the conduction band, $\varepsilon _{1} $ is the energy of the first subband due to space quantization along $z-$axis and $U(z)$ stands for the potential shape along z-axis, as in case of Fig. 1 or Fig. 3. In order to take into account the magnetic field arrangement ($B||e_{z} $, $e_{z} $is the unit vector of $z-$axis, and $B\bot j$, where $j$ is the current density) choose the next vector potential gauge: $A_{x} =By,$ $A_{y} =0,$ $A_{z} =0$.) Look again at Fig. 1 and Fig. 3; what is peculiar about these pictures is that regardless of the shape of QWs, rectangular or triangular, the structures are symmetric with respect to the centre of the barrier which is in the middle of the structure. The Hamiltonian, which describes the electron moving in the potential like that of Fig. 1 and Fig. 3, is invariant under inversion $z\to -z$ and hence, its eigenfunctions are double degenerate. However, due to the tunneling effect this degeneracy is removed and the lowest energy state in each quantum well becomes splitted into two, bounding and anti-bounding states represented by even and odd functions, respectively. We termed these states also as symmetric and anti-symmetric ones. Then the eigenfunctions of the Hamiltonian (A.1) are of the form:

\begin{equation}
\psi_{a(b)}(\vec{r})=\frac{1}{\sqrt
L_{x}}\exp(ik_{x}x)\chi_{m}(y)\varphi_{a(b)}(z).
\end{equation}

In the last expression $k_{x} $is the wave vector along $x-$axis, $\chi _{m} (y)$ is the wave function corresponding to the $m^{th}$ Landau level, $L_{x} $ is the structure length along $x-$axis, $\varphi _{a(b)} (z)$ and stands for the anti-symmetric (symmetric) state corresponding to the symmetric structure of the Fig. 1 or Fig. 3.

\appendix
\section*{Appendix B}
\setcounter{section}{2}

 Two electrons interacting in the $e-e$ scattering, are indistinguishable and hence, the two electron states (1,2) and ($1',2'$), before and after scattering event, should be described by the Hartree-Fock determinants as follows:

\begin{equation}
\psi_{1}\left(\vec{r_{1}}\right)\psi_{2}\left(\vec{r_{2}}\right)=\frac{1}{\sqrt{2}}
\left[\psi_{1}\left(\vec{r_{1}}\right)\psi_{2}\left(\vec{r_{2}}\right)-\psi_{1}
\left(\vec{r_{2}}\right)\psi_{2}\left(\vec{r_{1}}\right)\right].
\end{equation}

In order to calculate the matrix elements entering the expressions for $W_{\vec{k}\to \vec{k}'\, \left(\vec{k}'\to \vec{k}\right)}^{(a,b)} $ in (14) using the last expression and (A.2), one should substitute $\varphi _{1} (\vec{r}_{1} )$  $\varphi _{2} (\vec{r}_{2} )$, ($i,j=1,2$) by the next functions:

\begin{eqnarray}
&\psi_{jb}\left(\vec{r_{i}}\right)&=\frac{1}{\sqrt{L_{x}}}\exp\left(ik_{jx}x_{i}\right)\chi_{mj}\left(y_{i}\right)\varphi_{b}\left(z_{i}\right)=\nonumber\\&&
\frac{1}{\sqrt{L_{x}}}\varphi_{\vec{k}j}\left(x_{i}\right)\chi_{mj}\left(y_{i}\right)\varphi_{b}\left(z_{i}\right),\nonumber\\
&\psi_{ja}\left(\vec{r_{i}}\right)&=\frac{1}{\sqrt{L_{x}}}\exp\left(ik_{jx}x_{i}\right)\chi_{mj}\left(y_{i}\right)\varphi_{a}\left(z_{i}\right)=\nonumber\\&&
\frac{1}{\sqrt{L_{x}}}\varphi_{\vec{k}j}\left(x_{i}\right)\chi_{mj}\left(y_{i}\right)\varphi_{a}\left(z_{i}\right),
\end{eqnarray}

where $\varphi _{a} (z_{i} )$  and $\varphi _{b} (z_{i} )$ correspond to symmetric and anti-symmetric states, respectively. Here in the last expressions $k_{j} \equiv k_{jx} $- is the wave vector corresponding to the electron movement along $x-$axis perpendicular to $z-$axis. Then the corresponding matrix elements $M_{(1,2)\to (1',2')}^{(a),(b)} $ are of the form: 

\begin{eqnarray}
&M_{(1,2)\rightarrow
(1^{'},2^{'})}^{(a)}=\nonumber\\
						&\frac{1}{\sqrt{2L_{x}}}\int dV_{1} \int
            dV_{2}\varphi_{\vec{k^{'}}1}^{*}(x_{1})\chi_{m^{'}1}^{*}(y_{1})\varphi_{b}^{*}(z_{1})
            \varphi_{\vec{k^{'}}2}^{*}(x_{2})\chi_{m^{'}2}^{*}(y_{2})\varphi_{b}^{*}(z_{2})\times \nonumber\\
            &U^{ee}\varphi_{\vec{k}1}(x_{1})\chi_{m1}(y_{1})\varphi_{b}(z_{1})
            \varphi_{\vec{k}2}(x_{2})\chi_{m2}(y_{2})\varphi_{b}(z_{2})-\nonumber\\
            &\frac{1}{\sqrt{2L_{x}}}\int dV_{1} \int dV_{2}
            \varphi_{\vec{k^{'}}1}^{*}(x_{1})\chi_{m^{'}1}^{*}(y_{1})\varphi_{b}^{*}(z_{1})
            \varphi_{\vec{k^{'}}2}^{*}(x_{2})\chi_{m^{'}2}^{*}(y_{2})\varphi_{b}^{*}(z_{2})\times \nonumber\\
            &U^{ee}\varphi_{\vec{k}2}(x_{1})\chi_{m2}(y_{1})\varphi_{b}(z_{1})
            \varphi_{\vec{k}1}(x_{2})\chi_{m1}(y_{2})\varphi_{b}(z_{2}),\\
&M_{(1,2)\rightarrow
(1^{'},2^{'})}^{(b)}=\nonumber\\
						&\frac{1}{\sqrt{2L_{x}}}\int dV_{1} \int
            dV_{2}\varphi_{\vec{k^{'}}1}^{*}(x_{1})\chi_{m^{'}1}^{*}(y_{1})\varphi_{a}^{*}(z_{1})
            \varphi_{\vec{k^{'}}2}^{*}(x_{2})\chi_{m^{'}2}^{*}(y_{2})\varphi_{a}^{*}(z_{2})\times \nonumber\\
            &U^{ee}\varphi_{\vec{k}1}(x_{1})\chi_{m1}(y_{1})\varphi_{a}(z_{1})
            \varphi_{\vec{k}2}(x_{2})\chi_{m2}(y_{2})\varphi_{a}(z_{2})-\nonumber\\
            &\frac{1}{\sqrt{2L_{x}}}\int dV_{1} \int dV_{2}
            \varphi_{\vec{k^{'}}1}^{*}(x_{1})\chi_{m^{'}1}^{*}(y_{1})\varphi_{a}^{*}(z_{1})
            \varphi_{\vec{k^{'}}2}^{*}(x_{2})\chi_{m^{'}2}^{*}(y_{2})\varphi_{a}^{*}(z_{2})\times \nonumber\\
            &U^{ee}\varphi_{\vec{k}2}(x_{1})\chi_{m2}(y_{1})\varphi_{a}(z_{1})
            \varphi_{\vec{k}1}(x_{2})\chi_{m1}(y_{2})\varphi_{a}(z_{2}),
\end{eqnarray}
where $dV_{1} =dx_{1} dy_{1} dz_{1} $ and $dV_{2} =dx_{2} dy_{2} dz_{2} $. The first term in each of these formulae corresponds to the direct interaction, while the second one to the exchange interaction, and 

\begin{eqnarray}
&U^{ee}\equiv U^{ee} \left(x_{1},y_{1},z_{1},x_{2},y_{2},z_{2}\right)=\nonumber\\
&\frac{e^{2}}{\epsilon_{2DEG} \left(\vec{k}\right)
\sqrt{\left(x_{1}-x_{2}\right)^{2}+\left(y_{1}-y_{2}\right)^{2}+\left(z_{1}-z_{2}\right)^{2}}}.
\end{eqnarray}

Here $\varepsilon _{2DEG} \left(\vec{k}\right)$ is the dielectric constant of the two dimensional electron gas which incorporates all the screening effects in 2DEG. The integration in the formulae (B.3) and (B.4) is carrying out in such a way that at least one of the entries in denominator in (B.5) ($\left(x_{1} -x_{2} \right)^{2} $, $\left(y_{1} -y_{2} \right)^{2} $ or $\left(z_{1} -z_{2} \right)^{2} $) is not equal zero, guarantying the convergence of the integrals. This is the consequence of the fact that $\varphi _{jb} (\vec{r}_{i} )$, $\varphi _{ja} (\vec{r}_{i} )$ are the wave functions corresponding to some definite energy and this energy determines the minimum distance at which two electrons can approach each other.


\section*{References}
\begin{thebibliography}{10}
\bibitem {1} M. C. Bronsager, K. Flensberg, B. Yu-Kuang Hu, Antti-Pekka Jauho, {\it Phys. Rev. Lett.} {\bf 77}, 1366 (1996)
\bibitem {2} H. Rubel, A. Fischer, W. Dietsche, K. von Klitzing, K. Eberl, {\it Phys. Rev. Lett.} {\bf 78}, 1763 (1996)  
\bibitem {3} X. G. Feng, S. Zelakiewicz, H. Noh, T. J. Ragucci, T. J. Gramila, {\it Phys. Rev. Lett.} {\bf 81}, 3219 (1998)
\bibitem {4} J. G. S. Lok, S. Kraus, M. Pohlt, W. Dietsche, K. von Klitzing, W. Wegscheider, M. Bichler, {\it Phys. Rev. B.} {\bf 63}, 041305(R) (2001)
\bibitem {5} M. Kellog, I. B. Spielman, J. P. Eisenstein, L. N. Pfeiffer, and K. W. West, {\it Phys. Rev. Lett.} {\bf 88}, 126804 (2002)
\bibitem {6} H. S. Fresser, H. Frey, F. E. Prins, D. A. Haram, D. P. Kern, J. Bottcher and H. Kunzel, {\it Semicon. Sci. Technol.} {\bf 15}, 242 (2000) 
\bibitem {7} G. S. Boebinger, H. W. Jiang, L. N. Pfeiffer, and K. W. West,  {\it Phys. Rev. Lett.} {\bf 64}, 1793 (1990)
\bibitem {8} K. Ensslin, A. Wixforth, M. Sundaram, P. F. Hopkins, J. H. English, A. C. Gossard, {\it at. all.}, {\it Phys. Rev. B.} {\bf 47}, 1366 (1992).
\bibitem {9} K. M. Brown, N. Turner, J. T. Nicholls, E. H. Linfield, M. Pepper, D. A. Ritchie, G. A. C. Jones, {\it Phys. Rev. B} {\bf 50}, 15465 (1994)
\bibitem {10} Danhong Huang, M. O. Manasreh, {\it Phys. Rev. B} {\bf 54}, 2044 (1996)
\bibitem {11} J. P. Eisenstein, A. H. MacDonald, {\it Nature}, {\bf 432}, p.691, 9 December 2004
\bibitem {12} R. M Lewis, P. D. Ye, L. W. Engel, D. C. Tsui, L. N. Pfeiffer, K. W. West, {\it Phys. Rev. Lett.}, {\bf 89} 136804-1, 2002
\bibitem {13} R. M Lewis, Yong P. Chen,  L. W. Engel, D. C. Tsui,  L. N. Pfeiffer, K. W. West, {\it Phys. Rev. B.}, {\bf 71} 081301(R), 2005
\bibitem {14} G. Gervais, H. L. Stormer, D. C. Tsui, L. W. Engel, P. L. Kuhns, W. G. Moulton, A. P. Reyes, L. N. Pfeiffer, K. W. Baldwin, K. W. West, {\it Phys. Rev. B.}, {\bf 72}, 041310(R), 2005
\bibitem {15} L. A. Tracy, J. P. Eisenstein, L. N. Pfeiffer, K. W. West, {\it Phys. Cond. Mat.} 0511321, 2006
\bibitem {16} G. S. Boebinger, A. Passner, L. N. Pfeiffer and K. W. West,  {\it Phys. Rev. B.} {\bf 43}, 12673 (1991) 
\bibitem {17} T. Jungwarth, T. S. Lay, L. Smrcka, M. Shayegan, {\it Phys. Rev. B}, {\bf 56}, 1029 (1997)   
\bibitem {18} Y. W. Suen, J. Jo, M. B. Santos, L. W. Engel, S. W. Hwang and M. Shayegan, {\it Phys. Rev.B} {\bf 44}, 5947 (1991)
\bibitem {19} J. A. Simmons, S. K. Lyo, N. E. Harff and J. F. Klem, {\it Phys. Rev. Lett.} {\bf 73}, 2256 (1994)
\bibitem {20} T. Ihn, H. Carmona, P. C. Main, L. Eaves and M. Henini, {\it Phys. Rev.B} {\bf 54}, R2315 (1996) 
\bibitem {21} E. M. Sheregii, D. Ploch, M. Marchewka, G. Tomaka, A. Kolek, A. Stadler, K. Mleczko, D. Zak, W. Strupinski, A. Jasik, R. Jakiela, Inst. Physics Conf. Ser., No 187, Paper presented at 12th Int. Conf. Narrow Semiconductors, Toulouse, France, 3-7, July 2005, Taylor and Francis, New York, 2006, pp. 585- 590
\bibitem {22} M. Marchewka, E. M. Sheregii, D. Ploch, I. Tralle, G. Tomaka, M. Furdak, XVIII Int. Conf. High magnetic fields in the Semiconductor Physics, Wurtzbourg 2006, (accepted for publication)
\bibitem {23} E. M. Sheregii, D. Ploch, M. Marchewka, G. Tomaka, A. Kolek, A. Stadler, K. Mleczko, W. Strupinski, A. Jasik, R. Jakiela, Parallel magneto-transport in multiple quantum well structures, Fizika Nizkikh Temperatur (English version: Low Temperature Physics edited by the American Institute of Physics) 30, no.11, 2004 pp. 1146-1156
\bibitem {24} G. Tomaka, E. M. Sheregii, T. Kakol, W. Strupinski, R. Jakiela, A. Kolek, A. Stadler, K. Mleczko, {\it Crys. Res. Technol.} {\bf 38}, No. 3-5, 407-415 (2003) 
\bibitem {25} The measurements at the temperature of about of 4.2 K were performed at Rzeszow University of Technology. 
\bibitem {26} D. G. Seiler and A. E. Stephens, /in: Landau level Spectroscopy, vol. 27.2 of series Mod. Problems in Cond. Matt. Science, Eds: G.Landwehr and E. I. Rashba, eds of series: V. M. Agranovich, A. A. Maradudin, Nort-Holland, Amsterdam, pp. 1035- 1137 (1991) 
\bibitem {27} W. Zawadzki, J. Phys. C 16, 229 (1983)
\bibitem {28} J. H. Davies, The Physics of Low-dimensional Semiconductors, Cambr. Univ. Press, Cambr. (1997)
\bibitem {29} S. Datta, Electronic Transport in Mesoscopic System, Cambr. Univ. Press, Cambr. (1999) 
\bibitem {30} B. Gantmacher, Y. Levinson, Carrier scattering in metals and semiconductor, North-Holland, NY-Tokyo (1987).

\end{thebibliography}
\end{document}